\documentclass[reprint, longbibliography]{revtex4-1}

\usepackage[utf8]{inputenc}

\usepackage[lining,semibold]{libertine} 
\usepackage[libertine, cmintegrals, bigdelims, vvarbb]{newtxmath}

\usepackage{amsmath}
\usepackage{amsfonts}
\usepackage{mathrsfs}
\usepackage{gensymb}
\usepackage{bbm}
\usepackage{dsfont}

\usepackage{kbordermatrix}
\renewcommand{\kbldelim}{(}
\renewcommand{\kbrdelim}{)}

\usepackage{chemformula}

\usepackage{soul}
\usepackage{xcolor}

\makeatletter
\def\maketag@@@#1{\hbox{\m@th\normalfont\normalsize#1}}
\makeatother

\definecolor{webgreen}{rgb}{0,.5,0}
\definecolor{webbrown}{rgb}{.6,0,0}
\definecolor{grigio}{rgb}{.85,.85,.85} 
\definecolor{RoyalBlue}{rgb}{0.0, 0.14, 0.4}
\definecolor{skyblue1}{rgb}{0.45,0.62,0.81}
\definecolor{skyblue2}{rgb}{0.2,0.39,0.64}
\definecolor{skyblue3}{rgb}{0.13,0.29,0.53}
\definecolor{scarlet1}{rgb}{0.93,0.16,0.16}
\definecolor{scarlet2}{rgb}{0.8,0,0}
\definecolor{scarlet3}{rgb}{0.64,0,0}

\definecolor{g}{gray}{0.50}

\definecolor{keynotered}{rgb}{219,102,93}

\usepackage{hyperref}
\hypersetup{%
    colorlinks=true, linktocpage=true, pdfstartpage=1, pdfstartview=FitV,%
    breaklinks=true, pdfpagemode=UseNone, pageanchor=true, pdfpagemode=UseOutlines,%
    plainpages=false, bookmarksnumbered, bookmarksopen=true, bookmarksopenlevel=1,%
    hypertexnames=true, pdfhighlight=/O,
    urlcolor=webbrown, linkcolor=RoyalBlue, citecolor=webgreen, 
    pdftitle={},%
    pdfauthor={Francesco Avanzini},%
    pdfsubject={},%
    pdfkeywords={},%
    pdfcreator={pdfLaTeX},%
    pdfproducer={LaTeX REVTeX}%
}

\DeclareMathAlphabet{\mathpzc}{OT1}{pzc}{m}{it}

\newcommand{\crn}{CRN}
\newcommand{\effectivecycle}{pseudo-emergent}
\newcommand{\effectiverct}{effective}

\newcommand{\norm}[1]{\lVert#1\rVert}

\newcommand{\chemspecies}{\alpha}
\newcommand{\elrct}{\rho}

\newcommand{\cg}[1]{\hat{#1}}

\newcommand{\internalset}{X}
\newcommand{\internalspec}{x}
\newcommand{\chemostattedset}{Y}
\newcommand{\chemostattedspec}{y}

\newcommand{\fastset}{Q}
\newcommand{\fast}{q}
\newcommand{\slowset}{P}
\newcommand{\slow}{p}
\newcommand{\projslow}{\mathbb P}

\newcommand{\excurrent}{I}

\newcommand{\conslaw}{\lambda}
\newcommand{\slowconslaw}{\zeta}
\newcommand{\fastconslaw}{\xi}
\newcommand{\unbrokenconslaw}{{\zeta_{u}}}
\newcommand{\brokenconslaw}{{\zeta_{b}}}

\newcommand{\coefficient}{ f}

\newcommand{\fastcycle}{\gamma}
\newcommand{\internal}{\iota}
\newcommand{\effective}{\varepsilon}

\newcommand{\gelrct}{r}
\newcommand{\gcgrct}{c}

\newcommand{\brokenq}{\text{br}}
\newcommand{\breaker}{{p}}
\newcommand{\nobreaker}{{f}}
\newcommand{\refchempot}{Y_{\breaker}}

\newcommand{\matrixmoieties}{\mathbb M}
\newcommand{\moieties}{m}
\newcommand{\matrixconslaw}{\mathbb L}
\newcommand{\matrixconslawforces}{\mathbb{L}_{\chemostattedset_\nobreaker}^{\text{br}}}
\newcommand{\matrixconslawXun}{\mathbb{L}_\internalset^{\text{un}}}
\newcommand{\matrixconslawXbr}{\mathbb{L}_\internalset^{\text{br}}}

\newcommand{\stepone}{\text{1}}
\newcommand{\steptwo}{\text{2}}

\newcommand{\Aexpression}{\text{A,exp}}
\newcommand{\Bexpression}{\text{B,exp}}
\newcommand{\Adegradation}{\text{A,deg}}
\newcommand{\Bdegradation}{\text{B,deg}}

\begin{document}

\title{Thermodynamics of Non-Elementary Chemical Reaction Networks}
\newcommand\unilu{\affiliation{Complex Systems and Statistical Mechanics, Department of Physics and Materials Science, University of Luxembourg, L-1511 Luxembourg}}
\author{Francesco Avanzini}
 \email{francesco.avanzini@uni.lu}
\unilu
\author{Gianmaria Falasco}
 \email{gianmaria.falasco@uni.lu}
\unilu
\author{Massimiliano Esposito}
 \email{massimiliano.esposito@uni.lu}
\unilu

\date{\today}

\begin{abstract}
We develop a thermodynamic framework for closed and open chemical networks applicable to non-elementary reactions that do not need to obey mass action kinetics. 
It only requires the knowledge of the kinetics and of the standard chemical potentials, and makes use of the topological properties of the network (conservation laws and cycles).
Our approach is proven to be exact if the network results from a bigger network of elementary reactions where the fast-evolving species have been coarse grained.
Our work should be particularly relevant for energetic considerations in biosystems where the characterization of the elementary dynamics is seldomly achieved.
\end{abstract}

\maketitle


\section{Introduction}
Many processes in biology result from the combined effect of numerous elementary chemical reactions obeying mass-action kinetics. Well-known examples are enzymatic reactions, signaling cascades, and gene regulations. 
Since their detailed kinetic characterization at the elementary level is hard to achieve, they are often modeled with effective schemes obeying non-mass action kinetics, e.g., the Michaelis and Menten mechanism~\cite{Michaelis1913, Kenneth2011}, Hill functions~\cite{Hill1910} and many others~\cite{Cornish-Bowden2004}. 
These models can be thought of as resulting from a coarse graining of elementary reactions by eliminating fast-evolving species and their use is well established~\cite{Segel1989, Lee2009, Gunawardena2012, Gunawardena2014, Rao2003, Sinitsyn2009, Kim2014}.

However, characterizing the energetics at the level of these effective schemes remains an open challenge especially out-of-equilibrium.
Inspired by developments in stochastic thermodynamics~\cite{Jarzynski2011, Seifert2012, VanDenBroeck2015}, the nonequilibrium thermodynamics of open chemical networks of elementary reactions is nowadays well established \cite{Rao2016, Falasco2018a, Rao2018b}.
Recently, a thermodynamically consistent coarse graining strategy of the dynamics of biocatalysts was proposed in Ref.~\cite{Wachtel2018}. At steady-state this approach reproduces the correct dissipation at the coarse grained level.

Building on this approach, in the present work we develop a thermodynamics directly applicable to (closed and open) chemical networks of non-elementary reactions described by deterministic non-mass action rate equations.
Crucially, this approach can be validated using thermodynamics of elementary reactions. 
Indeed, if the non-elementary network can be constructed from a network of elementary reactions by coarse graining the fast evolving species, one can show that the thermodynamic quantities evaluated in both are equivalent.  

Our work is structured as follows. Starting from a network of elementary reactions and assuming time scale separation, we derive a thermodynamics for the network obtained after coarse graining the fast evolving species. We start by considering the dynamics of the closed network of the elementary reactions in Sec.~\ref{sec_elementary_dynamics} and of the coarse grained ones in Sec.~\ref{sec_coarse_grained_dynamics}. We then proceed by building the corresponding thermodynamics in Sec.~\ref{sec_thermo_closed}.
The dynamics of the open networks is considered in Sec.~\ref{sec_dynam_open} and its thermodynamics in Sec.~\ref{sec_thermo_open}. 
In Sec.~\ref{sec_final_example} we discuss how our previous findings can be used to characterize the thermodynamics of non-elementary networks when the full elementary description is not known. In Sec.~\ref{sec_conclusions} we summarize our results and discuss their implications.


\section{Elementary Dynamics of Closed \crn s\label{sec_elementary_dynamics}}

We consider here chemical reaction networks (\crn s) composed of chemical species $\boldsymbol \chemspecies = (\dots, \chemspecies, \dots)^\intercal$ undergoing elementary reactions~\cite{Svehla1993},
\begin{equation}
\boldsymbol \chemspecies\cdot \boldsymbol \nu_{+\elrct} \ch{<=>[ $k_{+\elrct}$ ][ $k_{-\elrct}$ ]} \boldsymbol \chemspecies\cdot \boldsymbol \nu_{-\elrct}\,,
\label{eq_elementary_reaction}
\end{equation}
with $\boldsymbol \nu_{\pm\elrct}$  (resp. $k_{\pm\elrct}$) the vector of the stoichiometric coefficients (resp. the kinetic constant) of the forward/backward reaction $\elrct$. The state of deterministic  \crn s is specified by the concentration vector $\boldsymbol z(t)=(\dots, [\chemspecies](t),\dots)^\intercal$ of all the chemical species $\boldsymbol \chemspecies$. For closed \crn s, its dynamics follows  the rate equation
\begin{equation}
\mathrm d_t \boldsymbol z(t)=\mathbb S \boldsymbol j(\boldsymbol z(t))\,,
\label{eq_dynamics_crns}
\end{equation}
where we introduce the stoichiometric matrix $\mathbb S$ and the current vector $ \boldsymbol j(\boldsymbol z)$. The stoichiometric matrix $\mathbb S$ codifies the topology of the network. Each $\elrct$ column ${\mathbb S}_\elrct$ of the stoichiometric matrix specifies the net variation of the number of molecules for each species undergoing the $\elrct$ elementary reaction~\eqref{eq_elementary_reaction}, ${\mathbb S}_\elrct =  \boldsymbol\nu_{-\elrct} - \boldsymbol\nu_{+\elrct} $. The current vector $\boldsymbol j(\boldsymbol z)=(\dots,j^\elrct(\boldsymbol z),\dots)^\intercal$ specifies the net reaction current for every $\elrct$ elementary reaction~\eqref{eq_elementary_reaction} as the difference between the forward $j^{+\elrct}(\boldsymbol z)$ and backward reaction current $j^{-\elrct}(\boldsymbol z)$
\begin{equation}
j^{\elrct}(\boldsymbol z)=j^{+\elrct}(\boldsymbol z) - j^{-\elrct}(\boldsymbol z)\,,
\end{equation}
with $j^{\pm\elrct}(\boldsymbol z)$ satisfying the mass-action kinetics~\cite{Groot1984,Laidler1987,Pekar2005},
\begin{equation}
 j^{\pm\elrct}(\boldsymbol z) = k_{\pm\elrct}\boldsymbol z^{\boldsymbol \nu_{\pm\elrct}}\,.
\end{equation}
Note that we use $\boldsymbol a^{\boldsymbol b} = \prod_ia_i^{b_i}$.
In the following, we will refer to Eq.~\eqref{eq_dynamics_crns} as the elementary dynamics of closed \crn s.

\paragraph*{Example.} In all the manuscript, we illustrate our findings using a modified version of the model discussed in Ref.~\cite{Sinitsyn2009}. It represents the transformation of two identical substrates \ch{S} into one product \ch{P}. The process is catalyzed by a membrane enzyme \ch{E} which interacts with the substrate after it is adsorbed \ch{S_m} by the membrane:
\begin{equation}
\begin{split}
\ch{S &<=>[ $k_{+1}$ ][ $k_{-1}$ ] S_m}\\
\ch{S_m + E &<=>[ $k_{+2}$ ][ $k_{-2}$ ] ES}\\
\ch{S_m + ES &<=>[ $k_{+3}$ ][ $k_{-3}$ ] ESS}\\
\ch{ESS &<=>[ $k_{+4}$ ][ $k_{-4}$ ] E + P}
\end{split}
\label{eq_crn_example}
\end{equation}
For this model, we introduced here the concentration vector and the current vector,
\begin{align}
  \boldsymbol{z} &= 
  \begin{pmatrix}
    [\ch{E}]	\\
    [\ch{ES}] \\
    [\ch{ESS}] \\
    [\ch{P}]	\\
    [\ch{S}]	\\
    [\ch{S_m}]	
  \end{pmatrix}
  \,,
 &
  \boldsymbol{j}(\boldsymbol{z}) &=
  \begin{pmatrix}
    k_{+1}[\ch{S}]-k_{-1}[\ch{S_m}]\\ 
    k_{+2}[\ch{S_m}][\ch{E}]-k_{-2}[\ch{ES}]\\ 
    k_{+3}[\ch{S_m}][\ch{ES}]-k_{-3}[\ch{ESS}]\\ 
    k_{+4}[\ch{ESS}]-k_{-4}[\ch{E}][\ch{P}]
  \end{pmatrix}
  \,,
\end{align}
as well as the stoichiometric matrix
\begin{equation}
\mathbb{S}=
 \kbordermatrix{
    & \color{g}1 &\color{g}2&\color{g}3&\color{g}4\cr
    \color{g}\ch{E}    & 0  & -1 & 0 & 1 \cr
    \color{g}\ch{ES}   & 0  & 1 & -1 & 0\cr
    \color{g}\ch{ESS}   &0 & 0 & 1 & -1 \cr
    \color{g}\ch{P}   & 0  & 0 & 0  & 1\cr
    \color{g}\ch{S} & -1  & 0 & 0 & 0\cr
    \color{g}\ch{S_m}   &1  & -1 & -1 & 0\cr
  }\,,
  \label{eq_stochiometric_matrix_example}
\end{equation}


\subsection{Topological Properties}
As shown in Ref.~\cite{Polettini2014,Rao2016}, the linear independent vectors $\{\boldsymbol \ell^\conslaw\}$ of the cokernel of the stoichiometric matrix 
\begin{equation}
\boldsymbol \ell^{\conslaw} \cdot \mathbb S=0
\label{eq_conservation_laws}
\end{equation}
are the conservation laws of Eq.~\eqref{eq_dynamics_crns}. Indeed, for each vector $\boldsymbol \ell^{\conslaw}$, the scalar $L^\conslaw(\boldsymbol z(t)) \equiv \boldsymbol \ell^{\conslaw}\cdot \boldsymbol z(t)$ is a conserved quantity, i.e., $\mathrm d_tL^\conslaw (\boldsymbol z(t))=  \boldsymbol \ell^{\conslaw} \cdot \mathbb S \boldsymbol j(\boldsymbol z(t)) =0 $. The linear independent vectors $\{\boldsymbol c_\internal\}$ of the kernel of the stoichiometric matrix 
\begin{equation}
\mathbb S \boldsymbol c_\internal=0
\label{eq_internal_cycles}
\end{equation}
are the internal cycles of Eq.~\eqref{eq_dynamics_crns}. They are sequence of reactions that leave the state of \crn s invariant. Any linear combination of internal cycles gives a steady-state current vector, i.e., $\mathbb S \overline{\boldsymbol j} = 0$ with $ \overline{\boldsymbol j} \equiv \boldsymbol c_\internal \psi^\internal$.
Note that in all the work  we use the Einstein notation: repeated upper-lower indices implies the summation over all the allowed values for the indices, i.e., $ \boldsymbol a_ib^i = \sum_i\boldsymbol a_i {b^i}$ and  $a_ib^i = \sum_i a_i {b^i}$.


\paragraph*{Example.} For the \crn~\eqref{eq_crn_example}, there are two conservation laws,
\begin{align}
  \boldsymbol{\ell}^{\ch{E}} &= \kbordermatrix{
    & \cr
    \color{g}\ch{E}	 	&1 \cr
    \color{g}\ch{ES} 	&1 \cr
    \color{g}\ch{ESS} 	&1 \cr
    \color{g}\ch{P} 		&0 \cr
    \color{g}\ch{S} 		&0 \cr
    \color{g}\ch{S_m} 	&0 \cr
  }\,&\text{ and }&
  &
  \boldsymbol{\ell}^{\ch{S}} &= \kbordermatrix{
    & \cr
    \color{g}\ch{E} 		&0 \cr
    \color{g}\ch{ES} 	&1 \cr
    \color{g}\ch{ESS} 	&2 \cr
    \color{g}\ch{P} 		&2 \cr
    \color{g}\ch{S} 		&1 \cr
    \color{g}\ch{S_m} 	&1 \cr
  }\,,
  \label{eq_conservation_law_example}
\end{align}
with a clear physical interpretation. The total concentration of the enzyme is given by $L^{\ch{E}}=\boldsymbol{\ell}^{\ch{E}} \cdot \boldsymbol z = [\ch{E}] + [\ch{ES}] + [\ch{ESS}]$, while the total concentration of the substrate is given by $L^{\ch{S}}=\boldsymbol{\ell}^{\ch{S}} \cdot \boldsymbol z =  [\ch{ES}] + 2 [\ch{ESS}] + 2 [\ch{P}] + [\ch{S}] + [\ch{S_m}]$.  The  \crn~\eqref{eq_crn_example} has no internal cycles.


\subsection{Equilibrium}
Closed \crn s must be \textit{detailed balanced}, namely the rate equation~\eqref{eq_dynamics_crns} admits an equilibrium steady-state $\boldsymbol z_{\text{eq}}$ characterized by vanishing reaction currents,
\begin{equation}
\boldsymbol j(\boldsymbol z_{\text{eq}})=0\,.
\end{equation}
This, together with mass-action kinetics, implies the so-called \textit{local detailed balance} condition for the kinetic constants $k_{\pm\elrct}$ of the chemical reactions
\begin{equation}
\frac{k_{+\elrct}}{k_{-\elrct}}=\boldsymbol z_{\text{eq}}^{\mathbb S_{\elrct}}\,.
\label{eq_local_detailed_balance_kinetic}
\end{equation}
It ensures that i) $\boldsymbol z_{\text{eq}}$ is the only steady-state of the rate equation~\eqref{eq_dynamics_crns} ii) $\boldsymbol z(t)$ relaxes to $\boldsymbol z_{\text{eq}}$ and hence the thermodynamic consistency~\cite{Polettini2014}.


\section{Coarse Grained Dynamics of Closed \crn s\label{sec_coarse_grained_dynamics}}
When the chemical species evolve over two different time scales, they can be divided into two disjoint subsets: the fast-evolving species $\fastset$ and the slow-evolving species $\slowset$~\cite{Gunawardena2012, Gunawardena2014}. 
 We apply the same splitting to the stoichiometric matrix 
\begin{equation}
\mathbb{S}=
\begin{pmatrix}
\mathbb S^\fastset \\ 
\mathbb S^\slowset \\
\end{pmatrix}
\end{equation}
and to the concentration vector $\boldsymbol z=(\boldsymbol \fast, \boldsymbol \slow)$. This also allows us to split the rate equation~\eqref{eq_dynamics_crns} into 
\begin{align}
\mathrm d_t \boldsymbol \fast(t)=\mathbb S^\fastset \boldsymbol j(\boldsymbol \fast(t), \boldsymbol \slow(t))\label{eq_dynamics_fast}\,,\\
\mathrm d_t \boldsymbol \slow(t)=\mathbb S^\slowset \boldsymbol j(\boldsymbol \fast(t), \boldsymbol \slow(t))\,.\label{eq_dynamics_slow}
\end{align}

The coarse graining procedure provides a closed dynamical equation for the slow species only. It is based on two assumptions collectively called \textit{time scale separation hypothesis} or \textit{quasi-steady-state assumption}. 
 The first is the existence, for every concentration $\boldsymbol\slow$ of the slow species, of the concentration vector $\cg { \boldsymbol \fast}(\boldsymbol \slow)$ such that the current 
\begin{equation}
\cg{\boldsymbol j} (\boldsymbol \slow)\equiv \boldsymbol j(\cg{ \boldsymbol \fast}(\boldsymbol \slow), \boldsymbol \slow)
\label{eq_cg_dynamics}
\end{equation}
is a steady-state current of Eq.~\eqref{eq_dynamics_fast} or, equivalently, $\cg{\boldsymbol j} (\boldsymbol\slow)\in \ker\mathbb S^\fastset$. This is a topological property of the \crn. The second is the equivalence, during the evolution of $\boldsymbol z(t)=(\boldsymbol\fast(t), \boldsymbol\slow(t))$ according to the rates equations~\eqref{eq_dynamics_fast} and~\eqref{eq_dynamics_slow}, between the actual concentration vector of the fast species $\boldsymbol\fast(t)$ and the steady-state one, i.e., $\boldsymbol\fast(t) = \cg{\boldsymbol\fast}(\boldsymbol \slow(t))$. This means physically that the concentrations of the fast species relax instantaneously to the steady-state corresponding to the frozen values of $\boldsymbol \slow(t)$. This occurs when i) the reactions involving only the $\slowset$ species are much slower than the reactions involving only $\fastset$ species and ii) the abundances of the $\slowset$ species are much higher than the abundances of the $\fastset$ species. This latter condition ensures that when the $\slowset$ and $\fastset$ species are coupled by a reaction, on the same time scale the concentrations of the $\fastset$ species can dramatically change, the concentrations of the $\slowset$ species remain almost constant.

Hence, ${\boldsymbol j(\boldsymbol \fast(t), \boldsymbol \slow(t))}$ in~\eqref{eq_dynamics_slow} becomes the steady-state current $\cg{\boldsymbol j}(\boldsymbol \slow)$ of Eq.~\eqref{eq_dynamics_fast}. It can now be written as a linear combination of the right null eigenvectors  $\mathbb S^\fastset \boldsymbol c_\fastcycle=0$,
\begin{equation}
\cg{\boldsymbol j}(\boldsymbol \slow) = \boldsymbol c_\fastcycle \psi^\fastcycle(\boldsymbol \slow)\,,
\label{eq_ss_current_vector}
\end{equation}
with $\boldsymbol \slow$ dependent coefficients $\{\psi^\fastcycle\}$. The specific expression of $\psi^\fastcycle(\boldsymbol \slow)$ is not discussed here and is not fundamental in what follows. When the dynamics of the fast species at fixed concentrations of the slow ones is linear, the problem is solved in Refs.~\cite{Gunawardena2012, Wachtel2018} using a diagrammatic method developed in Ref.~\cite{King1956, Hill1966}. Note that $\{\boldsymbol c_\fastcycle\}$ includes all the internal cycles $\{\boldsymbol c_\internal\}$ of Eq.~\eqref{eq_internal_cycles} and, in general, other vectors $\{\boldsymbol c_\effective\}$ called \effectivecycle\text{ }cycles \cite{Polettini2014,Rao2016,Wachtel2018}. The former characterize a sequence of reactions which upon completion leaves all species $(\boldsymbol\fast, \boldsymbol\slow)$ unchanged, while the latter only leaves the fast species unchanged. Each coefficient $ \psi^\fastcycle(\boldsymbol \slow)$ represents a current along the cycle $\fastcycle$.
By employing the steady-state current vector~\eqref{eq_ss_current_vector} in Eq.~\eqref{eq_dynamics_slow}, we obtain a closed dynamical equation for the slow species,
\begin{equation}
\mathrm d_t \boldsymbol \slow(t) =\mathbb S^\slowset \cg{\boldsymbol j}(\boldsymbol \slow(t))= \mathbb S^\slowset \boldsymbol c_\effective \psi^\effective(\boldsymbol \slow(t))\,,
\end{equation}
where we used $\mathbb S^\slowset \boldsymbol c_\internal=0$ because of Eq.~\eqref{eq_internal_cycles}. This coarse grained dynamics will be denoted as \textit{effective} for convenience. It can be rewritten in a more compact way,
\begin{equation}
\mathrm d_t \boldsymbol \slow(t) =\cg {\mathbb S} \boldsymbol \psi(\boldsymbol \slow(t))\,,
\label{eq_coarse_grained_dynamics}
\end{equation}
introducing the effective stoichiometric matrix $\cg {\mathbb S}$ and the effective current vector $\boldsymbol\psi(\boldsymbol \slow)$. The effective stoichiometric matrix $\cg {\mathbb S}$ codifies the net stoichiometry of the slow species along the \effectivecycle\text{ }cycles. Each $\effective$ column $\cg {\mathbb S}_\effective= \mathbb S^\slowset\boldsymbol c_\effective$ specifies the net variation of the number of molecules for each slow species along the $\effective$ \effectivecycle\text{ }cycle, i.e., the stoichiometry of the effective reactions. The effective current vector $\boldsymbol\psi(\boldsymbol \slow)$ collects the \effectivecycle\text{ }cycle currents $\boldsymbol \psi(\boldsymbol \slow) = (\dots,\psi^\effective(\boldsymbol \slow),\dots)^\intercal$. Unlike $\boldsymbol j(\boldsymbol z)$ in Eq.~\eqref{eq_dynamics_crns}, $\boldsymbol \psi(\boldsymbol \slow)$ does not, in general, satisfy mass-action kinetics. 

Note that, if there were no \effectivecycle\text{ }cycles, $(\cg{\boldsymbol\fast}(\boldsymbol \slow), \boldsymbol\slow)$ would be an equilibrium condition of the elementary dynamics~\eqref{eq_dynamics_crns}. The corresponding effective dynamics for the slow species would become trivial: $\mathrm d_t\boldsymbol\slow(t)=0$.

\begin{figure}[t]
  \centering
  \includegraphics[width=1.\columnwidth]{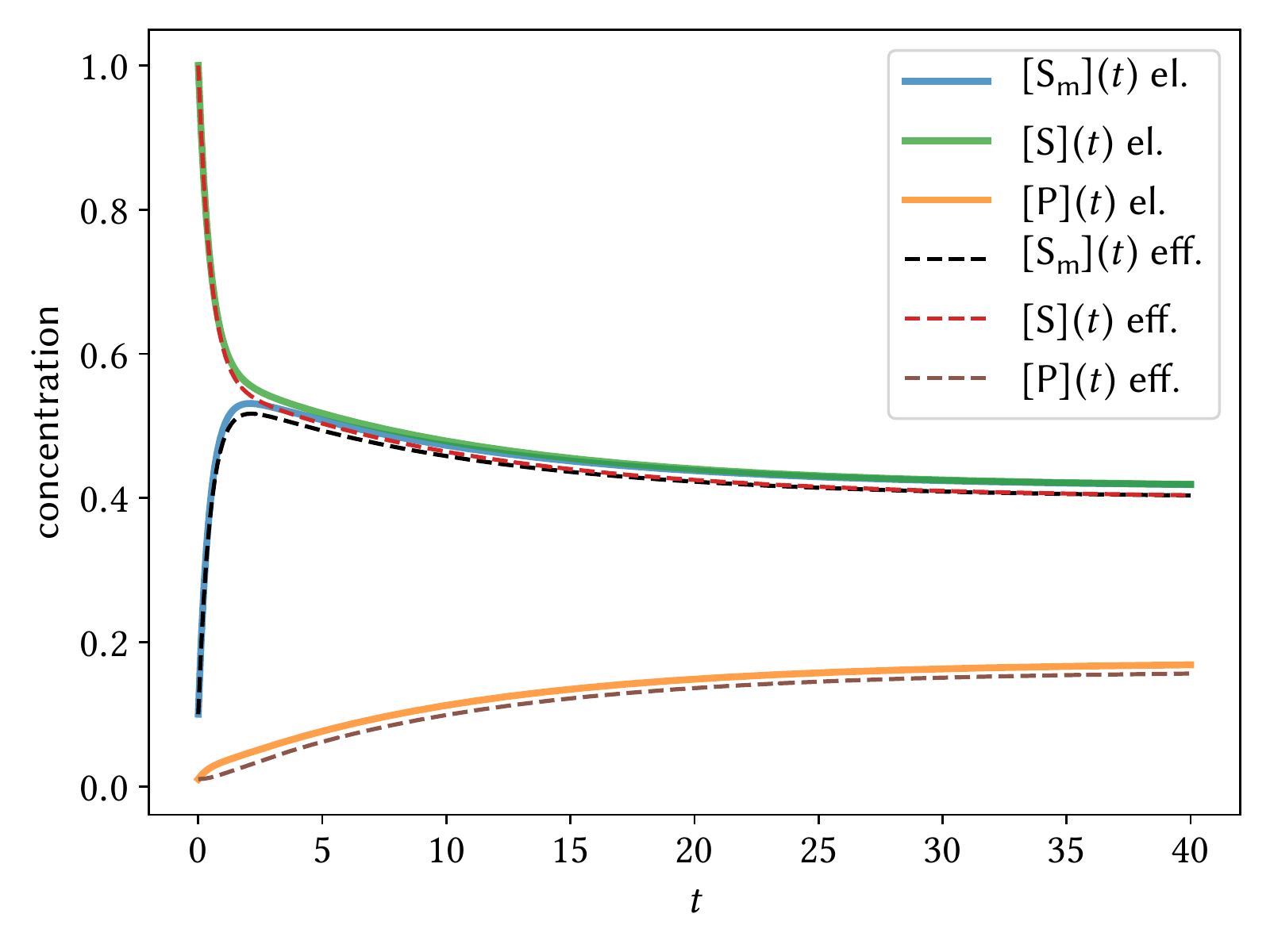}
  \caption{Elementary dynamics (el.) and effective dynamics (eff.) of the slow species of the \crn~\eqref{eq_crn_example}. Here, $k_{\pm\elrct}=1$ for every elementary reaction in~\eqref{eq_crn_example} and the initial condition $\boldsymbol z(0)$ is  $[\ch{E}](0)=0.1$, $[\ch{ES}](0)=0.05$, $[\ch{ESS}](0)=0.05$, $[\ch{P}](0)=0.01$, $[\ch{S}](0)=1$ and $[\ch{S_m}](0)=0.1$. We use $1/k_{+1}$ and $k_{+1}/k_{+2}$ as units of measure for time and concentration, respectively.}
  \label{fig_dynamics}
\end{figure}

\paragraph*{Example.} For the \crn~\eqref{eq_crn_example}, we split the chemical species as follows
\begin{align}
\underbrace{\left\{ \ch{E},\text{ } \ch{ES},\text{ } \ch{ESS}\right\}}_{\fastset} \cup \underbrace{\left\{\ch{P},\text{ } \ch{S},\text{ } \ch{S_m}\right\}}_{\slowset}\,.
\label{eq_fastslow_example}
\end{align}
Underlying this separation is the assumption that the concentration of the enzyme species changes much more quickly. Since there are no internal cycles of $\mathbb S$ of Eq.~\eqref{eq_stochiometric_matrix_example}, all the right null eigenvectors of $\mathbb S^\fastset$,
\begin{align}
  \boldsymbol{c}_{\text{ads}} &= 
  \kbordermatrix{
    & \cr
    \color{g}1& 1\cr
    \color{g}2& 0\cr  
    \color{g}3& 0\cr  
    \color{g}4& 0   
  }
  \,&\text{ and }&
 & \boldsymbol{c}_{\text{enz}} &= 
  \kbordermatrix{
    & \cr
    \color{g}1& 0\cr
    \color{g}2& 1\cr  
    \color{g}3& 1\cr  
    \color{g}4& 1     
  }\,,
\end{align}
are \effectivecycle\text{ }cycles. Thus, the effective stoichiometric matrix of Eq.~\eqref{eq_coarse_grained_dynamics} is given by
\begin{equation}
\cg{\mathbb{S}}=
 \kbordermatrix{
    & \color{g}\text{ads} &\color{g}\text{enz}\cr
    \color{g}\ch{P}   & 0  & 1 \cr
    \color{g}\ch{S} & -1  & 0 \cr
    \color{g}\ch{S_m}   &1  & -2 \cr
  }\,,
  \label{eq_effective_stochiometric_matrix_example}
\end{equation}
Each column of $\cg{\mathbb{S}}$ specifies the net stoichiometry of the following effective reactions for the slow species:
\begin{equation}
\begin{split}
\ch{S &<=>[ \text{ads} ][  ] S_m}\\
2\text{ }\ch{S_m &<=>[ \text{enz} ][  ] P}
\end{split}
\label{eq_cg_crn_example}
\end{equation}
In Fig.~\ref{fig_dynamics}, we compare the elementary dynamics of the \crn~\eqref{eq_crn_example} for the slow species and the effective one obtained by solving Eq.~\eqref{eq_dynamics_crns} and Eq.~\eqref{eq_coarse_grained_dynamics}, respectively. For this case, the effective current vector $\boldsymbol\psi(\boldsymbol\slow)$ is computed according to the procedure introduced in Refs.~\cite{King1956, Hill1966}:
\begin{align}
&\psi_\text{ads}(\boldsymbol\slow)=k_{+1}[\ch{S}]-k_{-1}[\ch{S_m}],\\
&\psi_\text{enz}(\boldsymbol\slow)=\frac{L^{\ch{E}}}{\mathcal D}\left\{ k_{+2}k_{+3}k_{+4}[\ch{S_m}]^2-k_{-2}k_{-3}k_{-4}[\ch{P}]\right\},
\end{align}
with $L^{\ch{E}}$ the total concentration of the enzyme and 
\begin{equation}
\begin{split}
\mathcal D=&k_{+2}k_{+3}[\ch{S_m}]^2+k_{+2}k_{+4}[\ch{S_m}]+k_{+2}k_{-3}[\ch{S_m}]+\\
&+k_{+3}k_{+4}[\ch{S_m}]+k_{-2}k_{-4}[\ch{P}]+k_{-3}k_{-4}[\ch{P}]\\
&+k_{+3}k_{-4}[\ch{S_m}][\ch{P}]+k_{-2}k_{-3}+k_{-2}k_{+4}\;.
\end{split}
\end{equation}


\subsection{Topological Properties\label{subs_topological_properties_cg}}

The conservation laws of the effective dynamics~\eqref{eq_coarse_grained_dynamics} are defined as the linear independent vectors of the cokernel of the effective stoichiometric matrix
\begin{equation}
\cg{\boldsymbol\ell}{}^{\slowconslaw}\cdot \cg{\mathbb S}=0\,,
\label{eq_definition_cg_conservation_laws}
\end{equation}
as in Eq.~\eqref{eq_conservation_laws}.  Indeed, each scalar $\cg{L}{}^\slowconslaw (\boldsymbol\slow(t))\equiv \cg{\boldsymbol\ell}{}^{\slowconslaw}\cdot \boldsymbol\slow(t)$ is a conserved quantity of the effective dynamics, i.e., $\mathrm d_t\cg{L}{}^\slowconslaw(\boldsymbol\slow(t)) =  \cg{\boldsymbol\ell}{}^{\slowconslaw}\cdot \cg{\mathbb S} \boldsymbol \psi(\boldsymbol \slow(t)) =0 $.

The conservation laws $\{\cg{\boldsymbol\ell}{}^{\slowconslaw}\}$ of \eqref{eq_definition_cg_conservation_laws} can be specified in terms of the conservation laws $\{\boldsymbol \ell^\conslaw\}$ of Eq.~\eqref{eq_conservation_laws} as follows.
First, we split $\{\boldsymbol \ell^\conslaw\}$ into two disjoint subsets: the conservation laws with null entries for the slow species $\{\boldsymbol\ell{}^{\fastconslaw}\}$ and the other conservation laws $\{\boldsymbol\ell{}^{\slowconslaw}\}$. Second, we consider the projection operator $\projslow$ onto the space of the slow species: $\projslow\boldsymbol z=\boldsymbol\slow$. Third, we identify each conservation law $\cg{\boldsymbol\ell}{}^{\slowconslaw}$ of Eq.~\eqref{eq_definition_cg_conservation_laws} as follows
\begin{equation}
\cg{\boldsymbol\ell}{}^{\slowconslaw}=\projslow\boldsymbol\ell{}^{\slowconslaw}\,.
\label{eq_entries_cg_conservation_laws}
\end{equation}
Indeed $\projslow{\boldsymbol\ell}{}^{\slowconslaw}\cdot\cg{\mathbb S}=0$. This can be verified considering that $\projslow{\boldsymbol\ell}{}^{\slowconslaw}\cdot\cg{\mathbb S}_\effective = \boldsymbol\ell{}^{\slowconslaw}\cdot\mathbb S\boldsymbol c_\effective=0$, where we only used the definition of \effectivecycle\text{ }cycle and conservation law, i.e., $\mathbb S^\fastset\boldsymbol c_\effective=0$ and $\boldsymbol\ell{}^{\slowconslaw}\cdot\mathbb S=0$.

Note that all the conservation laws of the effective stoichiometry matrix $\cg{\mathbb S}$ can be written as in Eq.~\eqref{eq_entries_cg_conservation_laws}. This follows from the rank nullity theorem for $\mathbb S$ and $\cg{\mathbb S}$, and the absence of cycles for  $\cg{\mathbb S}$. Indeed, suppose that there is a vector ${\boldsymbol \phi}\neq0$ such that $\cg{\mathbb S}{\boldsymbol \phi}=0$. This means that $\mathbb S^\slowset{\boldsymbol c_\effective \phi^\effective}=0$ and, consequently, $\boldsymbol c_\effective {\phi}^\effective$ is a right null eigenvector of both $\mathbb S^\slowset$ and $\mathbb S^\fastset$, i.e., an internal cycle. Since $\boldsymbol c_\effective {\phi}^\effective$ is a linear combination of only \effectivecycle\text{ }cycles $\{\boldsymbol c_\effective\}$, the effectively stoichiometric matrix cannot admit cycles.

\paragraph*{Example.} For the \crn~\eqref{eq_crn_example} and given the splitting between fast and slow species in~\eqref{eq_fastslow_example}, the only conservation law of $\mathbb S$ in~\eqref{eq_stochiometric_matrix_example} with no null entries for the slow species is $\boldsymbol \ell^{\ch{S}}$ of Eq.~\eqref{eq_conservation_law_example}. The corresponding conservation law of the effective dynamics is
\begin{equation}
 \cg{\boldsymbol{\ell}}{}^{\ch{S}}= \kbordermatrix{
    & \cr
    \color{g}\ch{P} 		&2 \cr
    \color{g}\ch{S} 		&1 \cr
    \color{g}\ch{S_m} 	&1 \cr
  }\,.
  \label{eq_cg_conservation_law_example}
\end{equation}
Given $\cg{\mathbb S}$ in Eq.~\eqref{eq_effective_stochiometric_matrix_example}, one easily verifies that $\cg{\boldsymbol{\ell}}{}^{\ch{S}}\cdot\cg{\mathbb S}=0$. The corresponding quantity $\cg{L}^{\ch S}=\cg{\boldsymbol{\ell}}{}^{\ch{S}}\cdot\boldsymbol\slow$ has the clear physical interpretation of the total concentration of substrate: $\cg{L}^{\ch S}= 2 [\ch{P}]+[\ch{S}]+[\ch{S_m}]$.

\subsection{Equilibrium}
The equilibrium condition $\boldsymbol \slow_{\text{eq}}$ of the effective dynamics~\eqref{eq_coarse_grained_dynamics} is defined by vanishing cycle currents,
\begin{equation}
\boldsymbol \psi(\boldsymbol \slow_{\text{eq}})=0\,.
\label{eq_cg_equilibrium}
\end{equation}
Its existence is granted by the \textit{local detailed balance} condition~\eqref{eq_local_detailed_balance_kinetic}. Indeed, the state $(\cg{\boldsymbol \fast}(\boldsymbol \slow_{\text{eq}}), \boldsymbol \slow_{\text{eq}})$ is an equilibrium of the elementary dynamics~\eqref{eq_dynamics_crns} by definition: it is a steady-state of the rate equation~\eqref{eq_dynamics_crns} which admits only equilibrium steady-states. 

Let us discuss now the equilibrium state to which the elementary and the effective dynamics relax given a unique initial condition $\boldsymbol z(0)=(\boldsymbol\fast(0), \boldsymbol\slow(0))$. The concentrations of the chemical species at equilibrium depend on the value of the conserved quantities. Therefore, to have similar concentrations for the slow species at equilibrium, also the values of $L{}^{\slowconslaw}(\boldsymbol z(0))=\boldsymbol l{}^{\slowconslaw}\cdot \boldsymbol z(0)$ and $\cg{L}^{\slowconslaw}(\boldsymbol \slow(0))=\cg{\boldsymbol l}{}^{\slowconslaw}\cdot \boldsymbol \slow(0)$ have to be similar. This means that the concentration of the fast species involved in the $\slowconslaw$ conservation laws must be much smaller than the concentration of the slow species. This is consistent with the \textit{time scale separation hypothesis} requiring much higher abundances for the slow-evolving than for the fast-evolving species.


\section{Thermodynamics of closed \crn s\label{sec_thermo_closed}}

We now derive the thermodynamics for the effective dynamics starting from the full elementary network. We emphasize that all the resulting thermodynamic quantities (e.g., entropy production and free energy) can be evaluated at the level of the effective dynamics, without any knowledge of the elementary dynamics. This is the key result of our work. 


\subsection{Local Detailed Balance}
The thermodynamic theory of \crn s presumes that all degrees of freedom other than concentrations are equilibrated at temperature $T$ and pressure of the solvent. In this way, thermodynamic state functions can be specified by their equilibrium form but expressed in terms of nonequilibrium  concentrations. The vector of chemical potentials is thus given by
\begin{equation}
\boldsymbol \mu(\boldsymbol z)=\boldsymbol{\mu}^{\circ}+RT\ln( \boldsymbol z)\,,
\end{equation}
with $\boldsymbol \mu^{\circ}$ the vector of the standard chemical potentials and $R$ the gas constant. The \textit{local detailed balance} condition~\eqref{eq_local_detailed_balance_kinetic} can be then restated to establish a correspondence between the elementary dynamics (i.e., kinetic constants $k_{\pm\elrct}$) and the thermodynamics (i.e., standard chemical potentials $\boldsymbol \mu^{\circ}$): 
\begin{equation}
\frac{k_{+\elrct}}{k_{-\elrct}}=\boldsymbol z_{\text{eq}}^{\mathbb S_{\elrct}}=\exp\left(-\frac{\boldsymbol\mu^{\circ}\cdot\mathbb S_\elrct}{RT}\right)\,.
\label{eq_local_detailed_balance}
\end{equation}

We now formulate the \textit{local detailed balance} condition for the effective dynamics~\eqref{eq_coarse_grained_dynamics}.
This is done by taking the product of the ratio $k_{+\elrct}/k_{-\elrct}$ along each \effectivecycle\text{ }cycle $\effective$
\begin{equation}
\prod_\elrct \left(\frac{k_{+\elrct}}{k_{-\elrct}}\right)^{c^\elrct_\effective}=\boldsymbol \slow_{\text{eq}}^{\cg{\mathbb S}_\effective}=\exp\left(-\frac{\cg{\boldsymbol\mu}^{\circ}\cdot\cg{\mathbb S}_\effective}{RT}\right)\,,
\label{eq_cg_local_detailed_balance}
\end{equation}
where $\cg{\boldsymbol\mu}^{\circ}=\projslow \boldsymbol\mu^{\circ}$ is the vector of the standard chemical potentials for the slow species $\slowset$. It is important to note that first Eq.~\eqref{eq_cg_local_detailed_balance} establishes the same correspondence between the equilibrium concentrations $\boldsymbol\slow_{\text{eq}}$ of the effective dynamics~\eqref{eq_coarse_grained_dynamics} and the standard chemical potentials $\cg{\boldsymbol\mu}^{\circ}$ as Eq.~\eqref{eq_local_detailed_balance}  does between $\boldsymbol z_{\text{eq}}$ and $\boldsymbol\mu^{\circ}$ for the elementary reactions. Second, the effective stoichiometric matrix $\cg{\mathbb S}$ plays the same role in Eq.~\eqref{eq_cg_local_detailed_balance} as $\mathbb S$ in Eq.~\eqref{eq_local_detailed_balance}. Third, the derivation of Eq.~\eqref{eq_cg_local_detailed_balance} employs only the topological properties of the \crn s and does not require the \textit{time scale separation hypothesis}.


\subsection{Gibbs Free Energy of Reaction\label{subsec_gibbs_free_energy_of_reaction}}
The Gibbs free energy of the $\elrct$ elementary reaction~\eqref{eq_elementary_reaction}, namely the thermodynamic force driving this reaction, is given by 
\begin{equation}
\Delta_\elrct G(\boldsymbol z)= \boldsymbol \mu(\boldsymbol z)\cdot\mathbb S_\elrct\,,
\label{eq_gibbs_free_energy_elementary_reaction}
\end{equation}
which becomes 
\begin{equation}
\Delta_\elrct G(\boldsymbol z)= - RT\ln\frac{j_{+\elrct}(\boldsymbol z)}{j_{-\elrct}(\boldsymbol z)}
\label{eq_gibbs_free_energy_elementary_reaction_currents}
\end{equation}
because of the \textit{local detailed balance}~\eqref{eq_local_detailed_balance}. At equilibrium $\Delta_\elrct G_{\text{eq}}\equiv \Delta_\elrct G(\boldsymbol z_{\text{eq}})=0$ since $j_{+\elrct}(\boldsymbol z_{\text{eq}})=j_{-\elrct}(\boldsymbol z_{\text{eq}})$. This means that ${\boldsymbol \mu}_{\text{eq}}$ belongs to the cokernel of ${\mathbb S}$.

We define now the Gibbs free energy of each effective reaction $\Delta_\effective G(\boldsymbol\slow)$.
 Collecting the Gibbs free energies of the elementary reactions in the vector $\Delta_\gelrct\boldsymbol G \equiv (\dots, \Delta_\elrct G, \dots)^\intercal$, $\Delta_\effective G$ is given by 
\begin{equation}
\Delta_\effective G(\boldsymbol\slow)\equiv \Delta_\gelrct\boldsymbol G(\boldsymbol z) \cdot\boldsymbol c_\effective = \cg{\boldsymbol \mu}(\boldsymbol \slow)\cdot\cg{\mathbb S}_\effective\,.
\label{eq_gibbs_free_energy_effective_cycle}
\end{equation}
We used Eq.~\eqref{eq_gibbs_free_energy_elementary_reaction}, $\mathbb S^\fastset\boldsymbol c_\effective=0$ and introduced the vector of the chemical potential of the slow species
\begin{equation}
\cg{\boldsymbol \mu}(\boldsymbol \slow)=\cg{\boldsymbol\mu}^\circ+RT\ln\boldsymbol \slow=\projslow\boldsymbol\mu(\boldsymbol z)\,.
\end{equation}
The expression of $\Delta_\effective G$ in Eq.~\eqref{eq_gibbs_free_energy_effective_cycle} is formally the same as $\Delta_\elrct G$ in Eq.~\eqref{eq_gibbs_free_energy_elementary_reaction}. At equilibrium $\Delta_\effective G_{\text{eq}}\equiv \Delta_\effective G(\boldsymbol\slow_{\text{eq}})=0$, because of the \textit{local detailed balance} condition~\eqref{eq_cg_local_detailed_balance}. This means that $ \cg{\boldsymbol \mu}_{\text{eq}}$ belongs to cokernel of $\cg{\mathbb S}$. Unlike $\Delta_\elrct G(\boldsymbol z)$, there is no analytical correspondence between $\Delta_\effective G(\boldsymbol\slow)$ and $\psi^{\pm\effective}(\boldsymbol\slow)$ and, in general,
\begin{equation}
\Delta_\effective G(\boldsymbol\slow) \neq -RT\ln\frac{\psi^{+\effective}(\boldsymbol\slow)}{\psi^{-\effective}(\boldsymbol\slow)}\,.\label{eq_flux_force_relation}
\end{equation}
Here, $\psi^{\pm\effective}(\boldsymbol\slow)$ are two positive defined currents such that $\psi^{\effective}(\boldsymbol\slow) = \psi^{+\effective}(\boldsymbol\slow) - \psi^{-\effective}(\boldsymbol\slow)$.
This breaks the flux-force relation at the coarse grained level as was already stressed in Ref.~\cite{Wachtel2018}.

Finally, we note that, as the \textit{local detailed balance}, $\Delta_\effective G$ is defined using only topological properties of the \crn s.

\paragraph*{Example.} For the \crn~\eqref{eq_crn_example} and given the splitting between fast and slow species in~\eqref{eq_fastslow_example}, the Gibbs free energy of the effective reactions in Eq.~\eqref{eq_cg_crn_example} is specified as
\begin{equation}
\begin{split}
\Delta_{\text{ads}}G&=\mu_{\ch{S_m}}-\mu_{\ch{S}}\,,\\
\Delta_{\text{enz}}G&=\mu_{\ch{P}} - 2 \mu_{\ch{S_m}}\,.
\end{split}
\end{equation}


\subsection{Entropy Production Rate\label{subs_entropy_production_rate}}
The entropy production rate of the elementary dynamics reads
\begin{equation}
T\dot{\Sigma}(t)= -\Delta_\gelrct\boldsymbol G(\boldsymbol z(t))\cdot \boldsymbol j (\boldsymbol z(t))=  -\Delta_\elrct G(\boldsymbol z(t)) j^{\elrct}(\boldsymbol z(t))\geq0 \,.
\label{eq_entropy_production}
\end{equation}
It quantifies the dissipation of the relaxation towards equilibrium. 
Because of Eq.~\eqref{eq_gibbs_free_energy_elementary_reaction_currents}, the dissipation of each elementary reaction $\elrct$ is also non-negative, $-\Delta_\elrct G(\boldsymbol z(t)) j_{\elrct}(\boldsymbol z(t))\geq0$ (without summation over $\elrct$).

We define the entropy production rate of the effective dynamics $\cg{\dot{\Sigma}}$ using the expression in Eq.~\eqref{eq_entropy_production}, but evaluated along the effective trajectory $(\cg{\boldsymbol\fast}(\boldsymbol\slow(t)),\boldsymbol\slow(t))$. Using Eqs.~\eqref{eq_cg_dynamics},~\eqref{eq_ss_current_vector} and~\eqref{eq_gibbs_free_energy_effective_cycle}, we obtain that
\begin{equation}
T\cg{\dot{\Sigma}}(t)= -\Delta_\effective G(\boldsymbol\slow(t))\psi^\effective(\boldsymbol\slow(t)) =- \Delta_\gcgrct\boldsymbol G(\boldsymbol\slow(t))\cdot \boldsymbol \psi(\boldsymbol\slow(t))\geq0\,,
\label{eq_cg_entropy_production}
\end{equation}
where we collected the Gibbs free energies of the \effectiverct\text{ }reactions in the vector ${\Delta_\gcgrct\boldsymbol G} \equiv (\dots, \Delta_\effective G, \dots)^\intercal$. The effective entropy production rate $\cg{\dot{\Sigma}}(t)$ is still non-negative by definition: using Eq.~\eqref{eq_gibbs_free_energy_elementary_reaction_currents} one proves that $\dot{\Sigma}(t)\geq0$  for every $\boldsymbol j(\boldsymbol z)$ including $\cg{\boldsymbol j}(\boldsymbol\slow)$.
However, unlike the elementary dynamics, it is not granted that $-\Delta_\effective G(\boldsymbol\slow(t))\psi_\effective(\boldsymbol\slow(t))\geq 0$ for every \effectiverct\text{ }reaction $\effective$ due to the lack of a flux-force relation~\eqref{eq_flux_force_relation} at the coarse grained level. 

If the \textit{time scale separation hypothesis} holds, namely $\boldsymbol j(\boldsymbol z)=\cg{\boldsymbol j}(\boldsymbol\slow)$, the two entropy production rates coincide:
\begin{equation}
\cg{\dot{\Sigma}}(t)= {\dot{\Sigma}}(t)\,.
\label{eq_equivalence_entropy_production}
\end{equation} 
When it does not hold, no bound constrains one entropy production rate respect to the other and $\cg{\dot{\Sigma}}$ can be lower or greater than ${\dot{\Sigma}}$. 
This was also observed in Ref.~\cite{Esposito2015} for driven systems.

\begin{figure}[t]
  \centering
  \includegraphics[width=1\columnwidth]{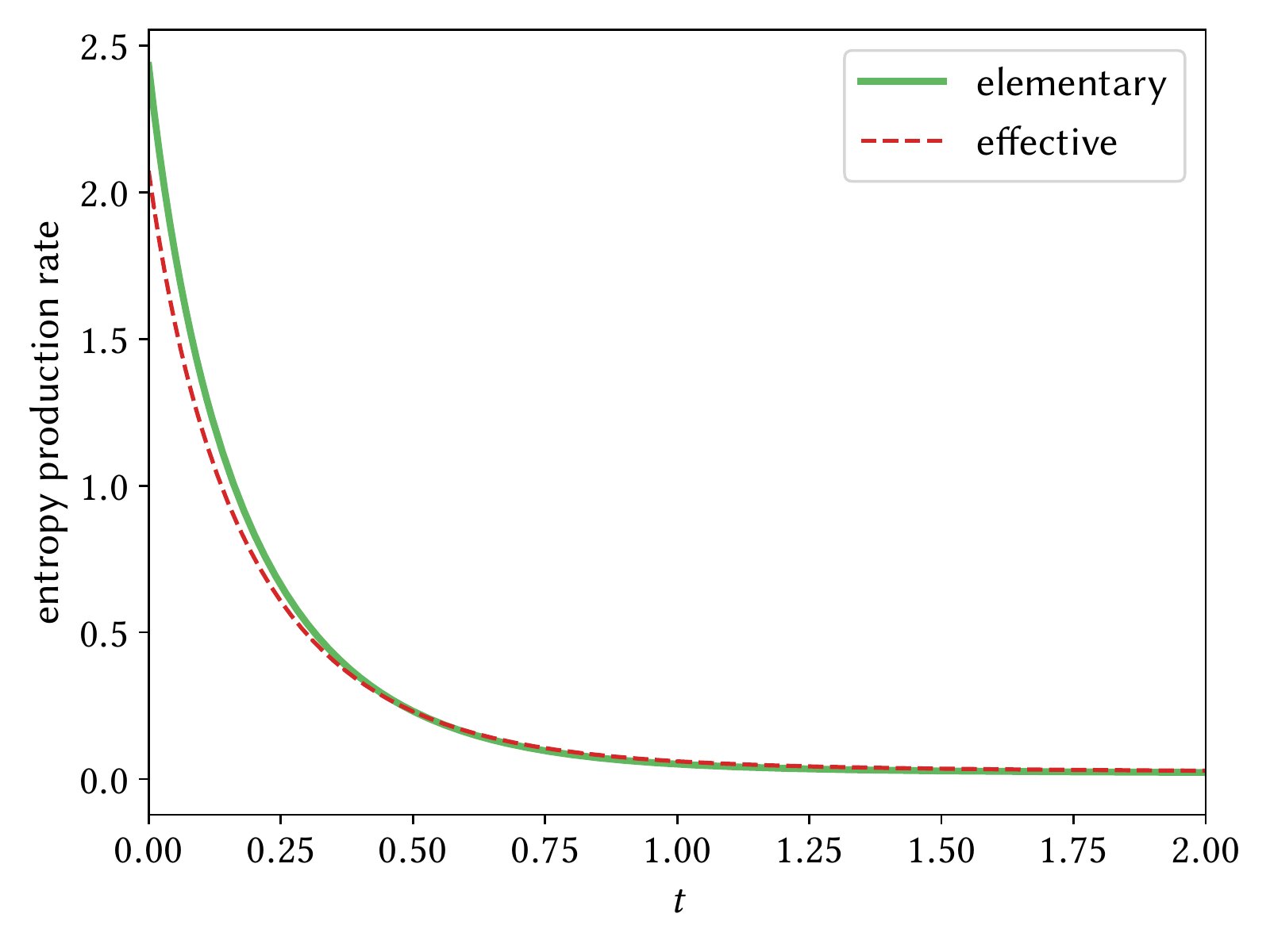}
  \caption{Entropy production rate of the elementary dynamics and effective dynamics in Fig.~\ref{fig_dynamics}. We use $RT(k_{+1})^2/k_{+2}$ as units of measure for the entropy production rate.}
  \label{fig_entropy}
\end{figure}

\paragraph*{Example.} For the \crn~\eqref{eq_crn_example} with the fast and slow species splitting~\eqref{eq_fastslow_example}, the entropy production rate of the elementary and of the effective dynamics are plotted in Fig.~\ref{fig_entropy}. While $\cg{\dot{\Sigma}}(t)<{\dot{\Sigma}}(t)$ for $t<0.5$, $\cg{\dot{\Sigma}}(t)>{\dot{\Sigma}}(t)$ for $t>0.5$.


\subsection{Gibbs Free Energy}
The Gibbs free energy of ideal dilute solution is given by
\begin{equation}
G(\boldsymbol z) =\boldsymbol\mu(\boldsymbol z)\cdot\boldsymbol z -RT \norm{\boldsymbol z}\,,
\label{eq_gibbs_free_energy}
\end{equation}
with $\norm{\boldsymbol a}\equiv \sum_i  a_i$. It is the proper thermodynamic potential of the elementary dynamics of closed \crn s since it has been proven in Ref.~\cite{Rao2016} that i) $\mathrm d_t G = -T\dot{\Sigma}\leq0$ and ii) $G\geq G_{\text{eq}}\equiv G(\boldsymbol z_{\text{eq}})$.

We now explain why the following Gibbs free energy is the proper thermodynamic potential of the effective dynamics:
\begin{equation}
\cg G(\boldsymbol \slow)=\cg{\boldsymbol\mu}(\boldsymbol \slow)\cdot\boldsymbol \slow -RT \norm{\boldsymbol \slow}\,.
\label{eq_cg_gibbs_free_energy}
\end{equation}

First, we notice that in general $\cg G(\boldsymbol \slow) \neq G(\boldsymbol z)$. 
This means that $\cg G$ does not estimate the exact free energy of \crn s. However, the time derivative of $\cg G$ according to the effective dynamics~\eqref{eq_coarse_grained_dynamics} reads
\begin{equation}\begin{split}
\mathrm d_t \cg G(\boldsymbol\slow(t)) &= \cg{\boldsymbol\mu}(\boldsymbol\slow(t))\cdot\cg{\mathbb S}\boldsymbol\psi(\boldsymbol\slow(t)) = {\Delta_\gcgrct\boldsymbol G}(\boldsymbol\slow(t))\cdot\boldsymbol \psi(\boldsymbol\slow(t))\\
	&=-T\cg{\dot{\Sigma}}(t)\leq0
\end{split}
\label{eq_dt_cg_gibbs_free_energy}
\end{equation}
where we used the definition of the Gibbs free energy of the \effectiverct\text{ }reactions given in Eq.~\eqref{eq_gibbs_free_energy_effective_cycle} and the entropy production rate~\eqref{eq_cg_entropy_production}. The variation of free energy in a time interval $[0,t]$ thus satisfies
\begin{equation}
\Delta \cg G(t)=-T\int_0^t\mathrm dt\text{ }\cg{\dot{\Sigma}}(t)=-T\int_0^t\mathrm dt\text{ }{\dot{\Sigma}}(t)=\Delta G(t)\,,
\label{eq_variation_free_energy_during_the_dynamics}
\end{equation}
as long as the \textit{time scale separation hypothesis} holds and, hence, $\cg{\dot{\Sigma}}(t)= {\dot{\Sigma}}(t)$. 

Second, one can prove that $\cg G(\boldsymbol p(t))\geq \cg G_{\text{eq}}\equiv \cg G(\boldsymbol\slow_{\text{eq}})$. Indeed, consider
\begin{equation}
\cg G_{\text{eq}} = \cg{\boldsymbol\mu}_{\text{eq}}\cdot\boldsymbol \slow_{\text{eq}} -RT \norm{\boldsymbol \slow_{\text{eq}}}\,.
\label{eq_cg_equilibrium_gibbs_free_energy}
\end{equation}
As mentioned in Subs.~\ref{subsec_gibbs_free_energy_of_reaction}, $\cg{\boldsymbol\mu}_{\text{eq}}$ belongs to the cokernel of $\cg{\mathbb S}$ and, therefore, it can be expressed as a linear combination of the conservation laws $\cg{\boldsymbol l}{}^{\slowconslaw}$:
\begin{equation}
\cg{\boldsymbol\mu}_{\text{eq}} = \coefficient_{\slowconslaw} \cg{\boldsymbol l}{}^{\slowconslaw}\label{eq_chem_potential_eq_cons_law}\,.
\end{equation}
Thus, the term $\cg{\boldsymbol\mu}_{\text{eq}}\cdot\boldsymbol \slow_{\text{eq}}$ in Eq.~\eqref{eq_cg_equilibrium_gibbs_free_energy} satisfies
\begin{equation}
\cg{\boldsymbol\mu}_{\text{eq}}\cdot\boldsymbol \slow_{\text{eq}} =  \coefficient_{\slowconslaw} \underbrace{\cg{\boldsymbol l}{}^{\slowconslaw}\cdot\boldsymbol \slow_{\text{eq}}}_{=\cg{L}{}^{\slowconslaw}}=\coefficient_{\slowconslaw} \cg{\boldsymbol l}{}^{\slowconslaw}\cdot\boldsymbol \slow(t)=\cg{\boldsymbol\mu}_{\text{eq}}\cdot\boldsymbol \slow(t)\,,
\end{equation}
where we used $\mathrm d_t\cg{L}{}^{\slowconslaw}=0$. This allows us to write the difference between $\cg G(\boldsymbol\slow(t))$ and $\cg G_{\text{eq}}$ as a relative entropy
\begin{equation}
\cg G(\boldsymbol\slow(t)) -\cg G_{\text{eq}} = RT\mathcal L (\boldsymbol \slow (t)\parallel \boldsymbol \slow_{\text{eq}})\,,
\label{eq_cg_thermo_potential}
\end{equation}
with
\begin{equation}
\mathcal L (\boldsymbol a\parallel \boldsymbol b)= \sum_ia_i\ln\left(\frac{a_i}{b_i}\right)-(a_i-b_i)\geq0\,,
\end{equation}
proving that $\cg G(\boldsymbol\slow (t))\geq \cg G_{\text{eq}}$. 

In summary, $\cg G$ in Eq.~\eqref{eq_cg_gibbs_free_energy} is the proper thermodynamic potential of the effective dynamics because of Eq.~\eqref{eq_dt_cg_gibbs_free_energy} and Eq.~\eqref{eq_cg_thermo_potential}. It also correctly quantifies the variation of the free energy of the whole \crn\text{ }because of Eq.~\eqref{eq_variation_free_energy_during_the_dynamics}.


\section{Dynamics of Open \crn s\label{sec_dynam_open}}
In open \crn s some species are exchanged with external reservoirs called chemostats.
We discuss in parallel the elementary dynamics and the effective one. 
We consider that the concentrations of chemostatted species are either fixed or slowly driven by the chemostats. 
These species are therefore treated as slow-evolving species.

We thus split $\slowset$ into two disjoint subsets:
the internal species $\internalset$ and the chemostatted species $\chemostattedset$. The former (as well as the fast species) evolve in time only because of the chemical reactions and their rate equation is unchanged. The latter evolve in time because of the chemical reactions and of matter flows with the chemostats. These are accounted in the rate equation by introducing the exchange current~$\boldsymbol \excurrent(t)$. For the elementary dynamics of the chemostatted species, the rate equation is given by
\begin{equation}
\mathrm d_t\boldsymbol \chemostattedspec(t)=\mathbb S^\chemostattedset\boldsymbol j (\boldsymbol z(t))+ \boldsymbol \excurrent(t)\,,
\label{eq_rate_equation_chemostatted_species}
\end{equation}
while for the effective dynamics it reads
\begin{equation}
\mathrm d_t\boldsymbol \chemostattedspec(t)=\cg{\mathbb S}^\chemostattedset\boldsymbol\psi(\boldsymbol\slow(t)) + \boldsymbol \excurrent(t)\,.
\label{eq_cg_rate_equation_chemostatted_species}
\end{equation}
Here we applied the splitting $\slowset=(\internalset,\chemostattedset)$ to the substoichiometric matrix $\mathbb S^\slowset$ and the effective stoichiometric matrix $\cg{\mathbb S}$,
\begin{align}
&\mathbb S^\slowset=\begin{pmatrix}
\mathbb S^\internalset \\ 
\mathbb S^\chemostattedset \\
\end{pmatrix}\,, 
&\cg{\mathbb S}=\begin{pmatrix}
\cg{\mathbb S}^\internalset \\ 
\cg{\mathbb S}^\chemostattedset \\
\end{pmatrix}\,, 
\label{eq_stoichiometric_matrix_internal}
\end{align}
as well as the vector $\boldsymbol\slow=(\boldsymbol\internalspec, \boldsymbol\chemostattedspec)$. 
Note that Eq.~\eqref{eq_rate_equation_chemostatted_species} and~\eqref{eq_cg_rate_equation_chemostatted_species} are merely definitions for the exchange current $\boldsymbol\excurrent(t)$. 
At the level of the effective dynamics, a slow variation of $\boldsymbol \chemostattedspec(t)$ corresponds to a slow variation of $\boldsymbol\excurrent(t)$.
Operationally, one may therefore equally well control $\boldsymbol\excurrent(t)$ and determine $\boldsymbol \chemostattedspec(t)$ through Eq.~\eqref{eq_cg_rate_equation_chemostatted_species}.
Both types of external control, on  $\boldsymbol \chemostattedspec(t)$ or $\boldsymbol\excurrent(t)$, can thus be considered.

\begin{figure}[t]
  \centering
  \includegraphics[width=1\columnwidth]{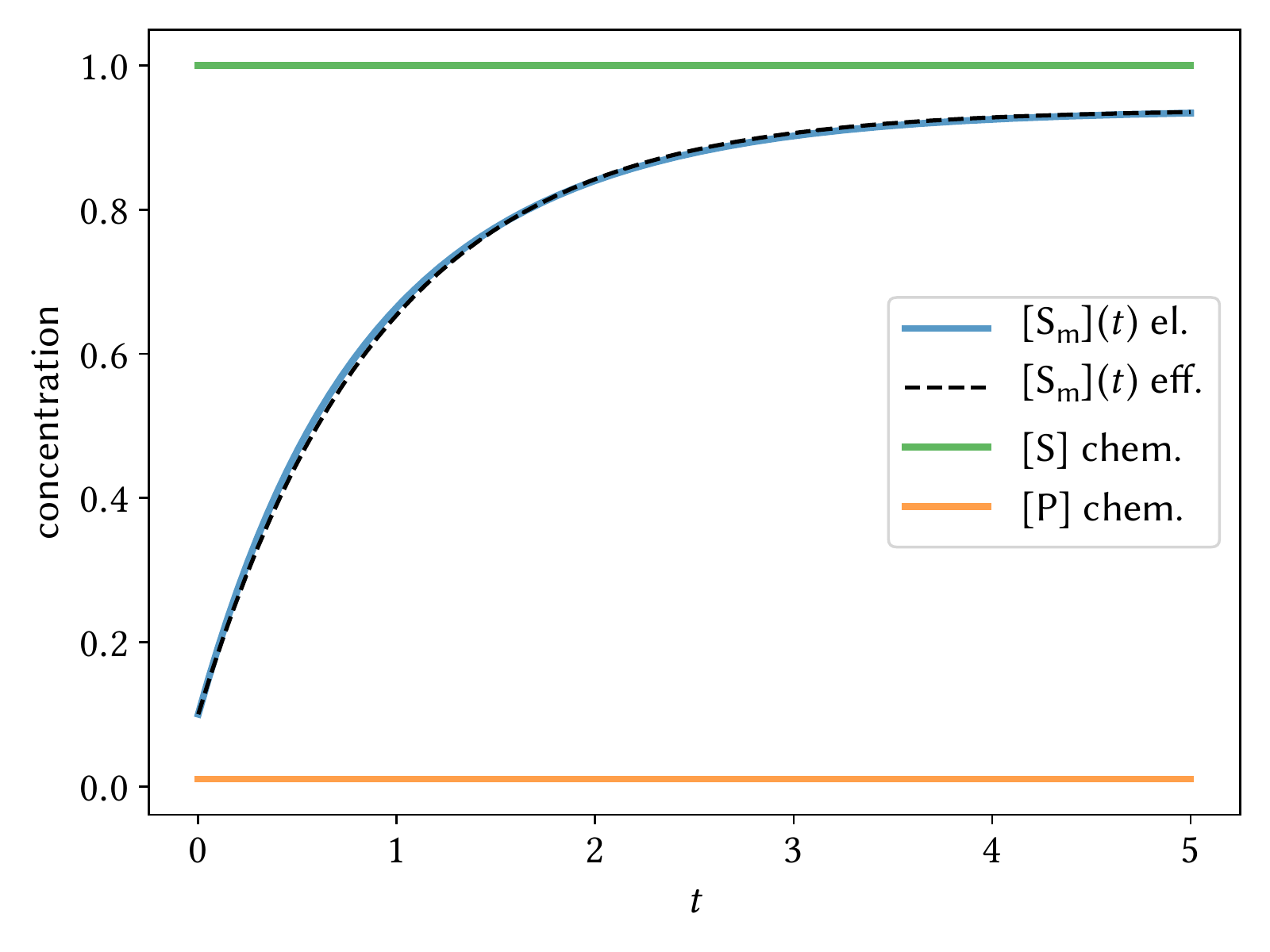}
  \caption{Elementary dynamics (el.) and effective dynamics (eff.) of the internal species $\ch{S_m}$ of the \crn~\eqref{eq_crn_example}, corresponding to Fig.~\ref{fig_dynamics}, when the concentrations $[\ch{S}]$ and $[\ch{P}]$ are kept constant by chemostats (chem.).}
  \label{fig_dynamics_open}
\end{figure}

\paragraph*{Example.} For the \crn~\eqref{eq_crn_example} and given the splitting between fast and slow species in~\eqref{eq_fastslow_example}, we chemostat the free substrate \ch{S} and the product \ch{P}. The adsorbed substrate \ch{S_{\text{m}}} is now the only internal species. Using the effective stoichiometric matrix in Eq.~\eqref{eq_effective_stochiometric_matrix_example}, we obtain the following two substoichiometric matrix 
\begin{align}
&\cg{\mathbb{S}}^\internalset=
 \kbordermatrix{
    & \color{g}\text{ads} &\color{g}\text{enz}\cr
    \color{g}\ch{S_m}   &1  & -2 \cr
  }
&\text{ and }&
&\cg{\mathbb{S}}^\chemostattedset=
 \kbordermatrix{
    & \color{g}\text{ads} &\color{g}\text{enz}\cr
    \color{g}\ch{P}   & 0  & 1 \cr
    \color{g}\ch{S} & -1  & 0 \cr
  }\,.
\label{eq_effective_stochiometric_matrix_internal_chemo_example}
\end{align}
In Fig.~\ref{fig_dynamics_open}, we compare the elementary dynamics of the \crn~\eqref{eq_crn_example} for the internal species and the effective one.


\subsection{Topological Properties\label{subs_broken_conservation_laws}}
When CRNs are open, some conservation laws do not correspond anymore to conserved quantities. Hence, we split the set of conservation laws into two disjoint subsets: the unbroken and the broken conservation laws.

The unbroken conservation laws are those with null entries for the chemostatted species. For the elementary dynamics, the conservation laws $\{\boldsymbol\ell^\fastconslaw\}$ of the fast species (see Subs.~\ref{subs_topological_properties_cg}) are unbroken by definition. Then, we represent with the vectors $\{\boldsymbol\ell^\unbrokenconslaw\}\subseteq\{\boldsymbol\ell^\slowconslaw\}$ the other unbroken conservation laws (if any). Indeed, the quantities $L^{\unbrokenconslaw}(\boldsymbol z(t))\equiv \boldsymbol \ell^{\unbrokenconslaw}\cdot\boldsymbol z(t)$, as well as $L^{\fastconslaw}(\boldsymbol z(t))\equiv \boldsymbol \ell^{\fastconslaw}\cdot\boldsymbol z(t)$, are conserved even if the \crn\text{ } is open: 
\begin{equation}
\mathrm d_tL^{\unbrokenconslaw}(\boldsymbol z(t))=\underbrace{{\boldsymbol \ell}^{\unbrokenconslaw}\cdot{\mathbb S}\boldsymbol j(\boldsymbol z(t))}_{=0}+\sum_{\chemspecies\in Y}\underbrace{{\ell}^{\unbrokenconslaw}_\chemspecies}_{=0} I^\chemspecies(t)=0\,.
\end{equation}
The broken conservation laws are all the other conservation laws, $\{\boldsymbol\ell^\brokenconslaw\}=\{\boldsymbol\ell^\slowconslaw\}\setminus\{\boldsymbol\ell^\unbrokenconslaw\}$. The corresponding quantities $L^\brokenconslaw(\boldsymbol z(t))= \boldsymbol\ell^\brokenconslaw\cdot\boldsymbol z(t)$ are not conserved:
\begin{equation}
\mathrm d_tL^{\brokenconslaw}(\boldsymbol z(t))=\underbrace{{\boldsymbol \ell}^{\brokenconslaw}\cdot{\mathbb S}\boldsymbol j(\boldsymbol z(t))}_{=0}+\sum_{\chemspecies\in Y}\underbrace{{\ell}^{\brokenconslaw}_\chemspecies}_{\neq0} I^\chemspecies(t)\neq0\,.
\end{equation}

Because of the correspondence in Eq.~\eqref{eq_entries_cg_conservation_laws}, the unbroken/broken conservation laws of  the effective dynamics are given, respectively, by
\begin{align}
&\cg{\boldsymbol\ell}{}^\unbrokenconslaw=\projslow\boldsymbol\ell^\unbrokenconslaw
&\text{ and }&
&\cg{\boldsymbol\ell}{}^\brokenconslaw=\projslow\boldsymbol\ell^\brokenconslaw\,.
\end{align}

Chemostatting a species does not always break a conservation law~\cite{Rao2016, Falasco2018a, Rao2018b, Rao2018a}. We thus distinguish the set of controlled species $Y_\breaker\subseteq Y$ breaking the conservation laws from the others $Y_\nobreaker= Y \setminus Y_\breaker$. Note that the number of $Y_\breaker$ species is equal to the number of broken conservation laws.
This allows us to introduce the so-called moieties. They represent parts of (or entire) molecules which are exchanged with the environment through the chemostats. For the elementary dynamics, their concentration vector is specified as
\begin{equation}
\boldsymbol \moieties(\boldsymbol z(t))\equiv\matrixmoieties^{-1}\boldsymbol L_{\brokenq}(\boldsymbol z(t))\,, 
\label{eq_moieties}
\end{equation}
while for the effective dynamics it is given by 
\begin{equation}
\cg{\boldsymbol \moieties}(\boldsymbol \slow(t))\equiv\matrixmoieties^{-1}\cg{\boldsymbol L}{}_{\brokenq}(\boldsymbol \slow(t))\,.
\label{eq_cg_moieties}
\end{equation}
We introduced the vectors of broken conserved quantities $\boldsymbol L_{\brokenq}(\boldsymbol z(t))=(\dots,\boldsymbol\ell^\brokenconslaw\cdot\boldsymbol z(t),\dots)^\intercal$ and $\cg{\boldsymbol L}_{\brokenq}(\boldsymbol \slow(t))=(\dots,\cg{\boldsymbol\ell}{}^\brokenconslaw\cdot\boldsymbol \slow(t),\dots)^\intercal$, and the matrix $\matrixmoieties$ with entries $\{\ell^{\brokenconslaw}_{\chemspecies}\}_{\chemspecies\in Y_\breaker}$ (see Ref.~\cite{Rao2016, Falasco2018a, Rao2018b} for details). The matrix  $\matrixmoieties$ is square and nonsingular, and it can be inverted obtaining $\matrixmoieties^{-1}$. Comparing Eq.~\eqref{eq_moieties} and Eq.~\eqref{eq_cg_moieties}, one notices that ${\boldsymbol \moieties}(\boldsymbol z(t))=\cg{\boldsymbol \moieties}(\boldsymbol \slow(t))$ if ${L}^{\brokenconslaw}(\boldsymbol z(t))=\cg{L}{}^{\brokenconslaw}(\boldsymbol \slow(t))$ for every broken conservation law, i.e., the slow-evolving species are much more abundant than the fast-evolving species. We stress finally that the number of moieties is equal to the number of $Y_\breaker$ species.

\paragraph*{Example.} For the \crn~\eqref{eq_crn_example} and given the splitting between fast and slow species in~\eqref{eq_fastslow_example}, the conservation law $\cg{\boldsymbol\ell}{}^{\ch{S}}$ in Eq.~\eqref{eq_cg_conservation_law_example} of the effective dynamics is broken when the free product \ch{P} is chemostatted. If we now chemostat the free substrate \ch{S}, no conservation law is broken. According to this sequence of chemostatting, the species \ch{P} belongs to the set $Y_{\breaker}$ and the matrix $\matrixmoieties$ is simply the scalar
\begin{equation}
\matrixmoieties = 2\,.
\end{equation}
The corresponding moiety, 
\begin{equation}
\moieties = [\ch{P}] + \frac{[\ch{S}]}{2} + \frac{[\ch{S_m}]}{2}\,,
\end{equation}
represents the total concentration of product exchanged between the two chemostats.


\section{Thermodynamics of Open \crn s\label{sec_thermo_open}}
We consider now the thermodynamic description of open \crn s. The semigrand Gibbs free energy
\begin{equation}
\mathcal G(\boldsymbol z)=G(\boldsymbol z)-\boldsymbol\mu_{\refchempot}(t)\cdot \boldsymbol \moieties(\boldsymbol z)\,,
\label{eq_semigrand_gibbs_free_energy}
\end{equation}
represents the proper thermodynamic potential for the elementary dynamics~\cite{Rao2016, Falasco2018a} since i) $\mathrm d_t\mathcal G =-T\dot{\Sigma}\leq0$ for autonomous detailed balanced systems and ii) $\mathcal G\geq\mathcal G_{\text{eq}}=\mathcal G(\boldsymbol z_{\text{eq}})$.
In analogy to equilibrium thermodynamics, $\mathcal G$ is defined from the Gibbs free energy~\eqref{eq_gibbs_free_energy} by eliminating the energetic contributions of the matter exchanged with the reservoirs. The latter amounts to concentration of the moieties  $\boldsymbol \moieties(\boldsymbol z)$ of Eq.~\eqref{eq_moieties}, times the reference values of their chemical potentials $\boldsymbol \mu_{\refchempot}(t)$ which is the vector collecting the values of chemical potentials fixed by the chemostats $Y_\breaker$. Because of the elementary dynamics~\eqref{eq_dynamics_crns} and taking into account the exchange current $\boldsymbol I(t)$ through Eq.~\eqref{eq_rate_equation_chemostatted_species}, the evolution of $\mathcal G(\boldsymbol z)$ is given by
\begin{equation}
\mathrm d_t \mathcal G(\boldsymbol z(t))=-T\dot{\Sigma}(t)+\dot{w}_{\text{driv}}(t)+\dot{w}_{\text{nc}}(t)\,.
\end{equation}
The entropy production rate $\dot{\Sigma}$ is specified in Eq.~\eqref{eq_entropy_production}. The the driving work rate $\dot{w}_{\text{driv}}$ accounts for the time dependent manipulation of the chemical potential of the $Y_\breaker$ chemostats,
\begin{equation}
\dot{w}_{\text{driv}}(t) = -(\mathrm d_t\boldsymbol\mu_{\refchempot}(t))\cdot\boldsymbol \moieties(\boldsymbol z(t))\,.
\end{equation}
The nonconservative work rate $\dot{w}_{\text{nc}}$ quantifies the energetic cost of sustaining fluxes of chemical species among the chemostats,
\begin{equation}
\dot{w}_{\text{nc}}(t) =\boldsymbol{\mathcal F}(t)\cdot\boldsymbol I(t)\,,
\label{eq_nc_work}
\end{equation}
by means of the force $\boldsymbol{\mathcal F}(t)=(\dots, \mu_\chemspecies(t)-(\boldsymbol\mu_{\refchempot}(t)\cdot\matrixmoieties^{-1})_{\brokenconslaw}l^{\brokenconslaw}_\chemspecies, \dots)_{\chemspecies\in Y}^\intercal$. In other words, this is the force keeping the system out of equilibrium (see also App.~\ref{app_reference_chemical_potential}).

For the effective dynamics, we introduce the following semigrand Gibbs free energy:
\begin{equation}
\cg{\mathcal G}(\boldsymbol\slow)=\cg G(\boldsymbol\slow)-\boldsymbol\mu_{\refchempot}(t)\cdot\cg{\boldsymbol \moieties}(\boldsymbol\slow)\,,
\label{eq_cg_semigrand_gibbs_free_energy}
\end{equation}
with $\cg{\boldsymbol \moieties}(\boldsymbol \slow)$ given in Eq.~\eqref{eq_cg_moieties}. 
Note that $\boldsymbol\mu_{\refchempot}(t)$ in Eq.~\eqref{eq_semigrand_gibbs_free_energy} and in Eq.~\eqref{eq_cg_semigrand_gibbs_free_energy} are exactly the same since they are imposed by the same chemostats. 
We now show that $\cg{\mathcal G}(\boldsymbol\slow)$ is the proper thermodynamic potential of the effective dynamics of open \crn s.

In general $\cg{\mathcal G}(\boldsymbol \slow)\neq \mathcal G(\boldsymbol z)$ since $ \cg G(\boldsymbol \slow)\neq G(\boldsymbol z)$. 
However, we will now see that their variation in time can be very similar. 
According to the effective dynamics~\eqref{eq_coarse_grained_dynamics} and taking into account the exchange current $\boldsymbol I(t)$ through Eq.~\eqref{eq_cg_rate_equation_chemostatted_species}, the evolution of $\cg{\mathcal G}(\boldsymbol \slow)$ is given by 
\begin{equation}
\mathrm d_t \cg{\mathcal G}(\boldsymbol \slow(t))=-T\cg{\dot{\Sigma}}(t)+\cg{\dot{w}}_{\text{driv}}(t)+\cg{\dot{w}}_{\text{nc}}(t)\,.
\label{eq_cg_semigrand_gibbs_free_energy_dynamics}
\end{equation}
The entropy production rate $\cg{\dot{\Sigma}}$  of Eq.~\eqref{eq_cg_entropy_production} satisfies $\cg{\dot{\Sigma}}={\dot{\Sigma}}$ as long as the \textit{time scale separation hypothesis} holds (see Subs.~\ref{subs_entropy_production_rate}). The driving work rate at the effective level,
\begin{equation}
\cg{\dot{w}}_{\text{driv}}(t)= -(\mathrm d_t\boldsymbol\mu_{\refchempot}(t))\cdot\cg{\boldsymbol \moieties}(\boldsymbol\slow(t))\,,
\end{equation}
corresponds to $\dot{w}_{\text{driv}}$ of the elementary dynamics when $\cg{\boldsymbol \moieties}(\boldsymbol\slow(t))={\boldsymbol \moieties}(\boldsymbol z(t))$. This occurs when $\cg{L}^{\brokenconslaw}= {L}^{\brokenconslaw}$ (see Subs.~\ref{subs_broken_conservation_laws}). 
The nonconservative work rate $\cg{{\dot{w}}}_{\text{nc}}$ has the same expression as the one for the elementary dynamics specified in Eq.~\eqref{eq_nc_work}. However, the numerical values of $\cg{{\dot{w}}}_{\text{nc}}$ and ${{\dot{w}}}_{\text{nc}}$ can be different because different currents $\boldsymbol I(t)$ might be necessary to set the same concentrations of the chemostatted species for the elementary and effective dynamics if the  \textit{time scale separation hypothesis} does not hold perfectly. 
We can thus conclude that the variation of the semigrand Gibbs free energy in a time interval $[0,t]$ satisfies
\begin{equation}
\Delta\cg{\mathcal G} = \int_0^t\mathrm dt\text{ }\mathrm d_t \cg{\mathcal G}=\int_0^t\mathrm dt\text{ }\mathrm d_t {\mathcal G} = \Delta{\mathcal G}\,
\end{equation}
as long as the \textit{time scale separation hypothesis} holds (granting also $\cg{L}^{\brokenconslaw}= {L}^{\brokenconslaw}$).

If the open system is autonomous (no driving is performed, $\cg{\dot{w}}_{\text{driv}}=0$) and detailed balance (all the chemostatted species break a conservation law, $\cg{\dot{w}}_{\text{nc}}=0$), then $\mathrm d_t \cg{\mathcal G} = -T\cg{\dot{\Sigma}}\leq0$. 

Finally, we show that $\cg{\mathcal G}(\boldsymbol\slow(t))\geq \cg{\mathcal G}_{\text{eq}}\equiv \cg{\mathcal G}(\boldsymbol\slow_{\text{eq}})$, where the equilibrium state $\boldsymbol\slow_{\text{eq}}$ is set by the $Y_{\breaker}$ chemostats (see App.~\eqref{app_reference_chemical_potential}). To this aim, consider that $\cg{\boldsymbol\mu}_{\text{eq}}$ belongs to the cokernel of $\cg{\mathbb S}$ and can thus be expressed as a linear combination of the conservation laws:
\begin{equation}
\cg{\boldsymbol\mu}_{\text{eq}} =\coefficient_{\unbrokenconslaw} \cg{\boldsymbol \ell}{}^{\unbrokenconslaw}+ \coefficient_{\brokenconslaw} \cg{\boldsymbol \ell}{}^{\brokenconslaw}\,,
\end{equation}
where now we distinguish between unbroken and broken conservation laws.
Thus, $\cg{\boldsymbol\mu}_{\text{eq}}\cdot\boldsymbol \slow_{\text{eq}}$ satisfies now
\begin{equation}
\begin{split}
\cg{\boldsymbol\mu}_{\text{eq}}\cdot\boldsymbol \slow_{\text{eq}} &=  \coefficient_{\unbrokenconslaw} \underbrace{\cg{\boldsymbol \ell}{}^{\unbrokenconslaw}\cdot\boldsymbol \slow_{\text{eq}}}_{=\cg{L}{}^{\unbrokenconslaw}} + \coefficient_{\brokenconslaw} {\cg{\boldsymbol \ell}{}^{\brokenconslaw}\cdot\boldsymbol \slow_{\text{eq}}}\\
&= \cg{\boldsymbol\mu}_{\text{eq}}\cdot\boldsymbol \slow(t) -  \coefficient_{\brokenconslaw} {\cg{\boldsymbol \ell}{}^{\brokenconslaw}\cdot\boldsymbol \slow(t)}+  \coefficient_{\brokenconslaw} {\cg{\boldsymbol \ell}{}^{\brokenconslaw}\cdot\boldsymbol \slow_{\text{eq}}}\\
&= \cg{\boldsymbol\mu}_{\text{eq}}\cdot\boldsymbol \slow(t) - \boldsymbol \mu_{\refchempot}\cdot \cg{\boldsymbol \moieties}(\boldsymbol \slow(t))+\boldsymbol\mu_{\refchempot}\cdot\cg{\boldsymbol \moieties}_{\text{eq}}
\end{split}
\label{eq_finalstep_semigran_equilibrium}
\end{equation}
where we used $\mathrm d_t\cg{L}{}^{\unbrokenconslaw}=0$ and the last step is discussed in App.~\eqref{app_reference_chemical_potential}.  This allows us to write $\cg{\mathcal G}_{\text{eq}}$ as
\begin{equation}
\cg{\mathcal G}_{\text{eq}} = \cg{\boldsymbol\mu}_{\text{eq}}\cdot\boldsymbol \slow(t) - RT \norm{\boldsymbol \slow_{\text{eq}}}- \boldsymbol\mu_{\refchempot}\cdot \cg{\boldsymbol \moieties}(\boldsymbol \slow(t))
\end{equation}
and, consequently, the difference between $\cg{\mathcal G}(\boldsymbol \slow(t))$ and $\cg{\mathcal G}_{\text{eq}}$ as a relative entropy
\begin{equation}
\cg{\mathcal G}(\boldsymbol \slow(t)) - \cg{\mathcal G}_{\text{eq}} = RT\mathcal L (\boldsymbol \slow (t)\parallel \boldsymbol \slow_{\text{eq}}) \geq 0 \,.
\end{equation}
This proves that $\cg{\mathcal G}(\boldsymbol \slow(t))\geq \cg{\mathcal G}_{\text{eq}}$. 

\begin{figure}[t]
  \centering
  \includegraphics[width=1\columnwidth]{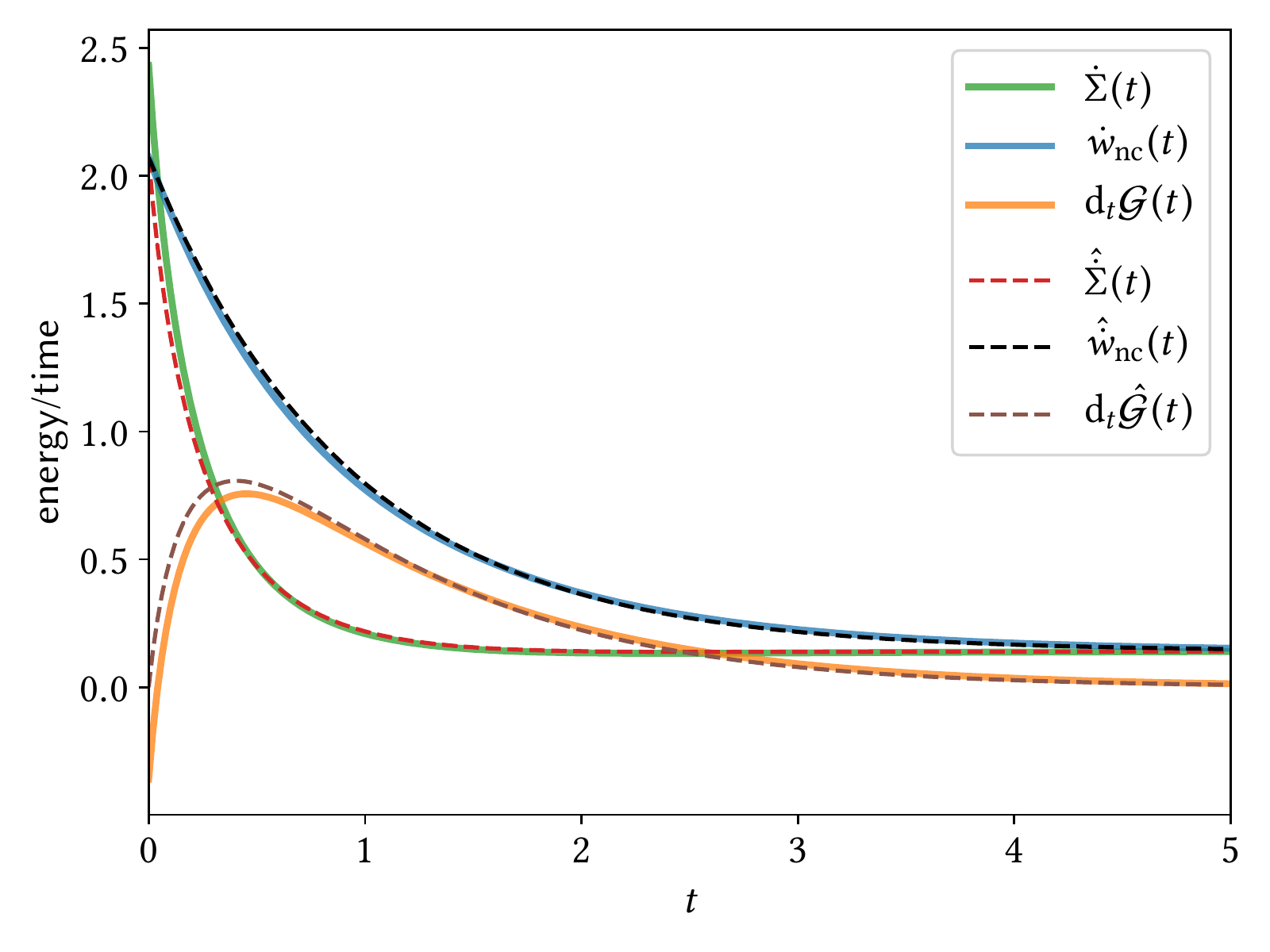}
  \caption{Thermodynamic quantities of the elementary dynamics and effective dynamics in Fig.~\ref{fig_dynamics_open}. Here, $\mu^{\circ}_{\ch{E}}=\mu^{\circ}_{\ch{S}}=\mu^{\circ}_{\ch{S_m}}=1$, $\mu^{\circ}_{\ch{ES}} = \mu^{\circ}_{\ch{P}}=2$ and $\mu^{\circ}_{\ch{ESS}}=3$. We use $RT(k_{+1})^2/k_{+2}$ as units of measure for the thermodynamic quantities.}
  \label{fig_thquant}
\end{figure}

\paragraph*{Example.} For the \crn~\eqref{eq_crn_example} and given the splitting between fast and slow species in~\eqref{eq_fastslow_example}, we compare the thermodynamic quantities of the elementary and of the effective dynamics in Fig.~\ref{fig_thquant}.



\section{Effective networks without elementary counterpart\label{sec_final_example}}
Our thermodynamic framework can be applied directly to effective networks even if the full elementary description is not available. 
Indeed, all the expressions (e.g., Eqs.~\eqref{eq_gibbs_free_energy_effective_cycle},~\eqref{eq_cg_entropy_production},~\eqref{eq_cg_gibbs_free_energy} and~\eqref{eq_cg_semigrand_gibbs_free_energy}) of the effective thermodynamic quantities require the knowledge of only the effective dynamics~\eqref{eq_coarse_grained_dynamics} and the standard chemical potentials $\cg{\boldsymbol\mu}^{\circ}$ of the slow species. This is the key result of our work. 

However, our framework should be applied only to effective models which are compatible with an underlying elementary network satisfying the \textit{time scale separation hypothesis}.
Indeed, we proved the thermodynamic consistency only in this case.
On the one hand, one should have some physical evidence supporting that the models result from elementary reactions satisfying the \textit{time scale separation hypothesis}.
On the other hand, the effective models must exhibit the following three properties to be thermodynamically consistent.

First, in absence of any chemostatted species (i.e., closed \crn s), the effective model must relax to an equilibrium~\eqref{eq_cg_equilibrium}. This is fundamental for energetic considerations. Physically, it means that every system has to reach an equilibrium when there are no energy sources (the chemostats) balancing the dissipation. Mathematically, it means that the effective reactions must be reversible and effective current vector $\boldsymbol \psi(\boldsymbol \slow)$ has to admit a nontrivial equilibrium steady-state, i.e., $\exists \boldsymbol \slow_{\text{eq}}\neq0$ such that $\boldsymbol \psi( \boldsymbol \slow_{\text{eq}})=0$. If the effective model relaxes to a nonequilibrium steady-state without chemostatted species, then it means that it does not account for hidden sources of energy and any energetic consideration becomes meaningless.
Second, provided that the \textit{time scale separation hypothesis} holds, there must be no cycles in the effective network as we proved in Subs.~\ref{subs_topological_properties_cg}.
Third, provided that the \textit{time scale separation hypothesis} holds, the effective entropy production rate~\eqref{eq_cg_entropy_production} must be greater than or equal to zero. The property $\cg{\dot{\Sigma}}(t)\geq0$ is not granted anymore if the cycle current vector $\boldsymbol \psi$ is not specified according to the coarse graining procedure discussed in Sec.~\ref{sec_coarse_grained_dynamics}. 
In this respect, an effective dynamics giving rise to $\cg{\dot{\Sigma}}(t)< 0$ is necessarily thermodynamically inconsistent.

We now consider two examples, one that can and the other that cannot be characterized using our framework.

\subsection{Example 1}

Consider the following reactions
\begin{equation}
\ch{S <=>[ $\psi_\stepone$ ][  ] I <=>[ $\psi_\steptwo$ ][  ] P},
\label{eq_model_meta}
\end{equation}
satisfying the dynamics 
\begin{equation}
\left\{
\begin{split}
&\mathrm d_t  [\ch{S}](t)= -\psi_\stepone( [\ch{S}](t),  [\ch{I}](t))\\
&\mathrm d_t  [\ch{I}](t)=\psi_\stepone( [\ch{S}](t),  [\ch{I}](t))-\psi_\steptwo(  [\ch{I}](t), [\ch{P}](t))\\
&\mathrm d_t  [\ch{P}](t)=\psi_\steptwo(  [\ch{I}](t), [\ch{P}](t))
\end{split}
\right.
\label{eq_ds_meta}
\end{equation}
with
\begin{equation}
\footnotesize
\begin{split}
&\psi_\stepone([\ch{S}], [\ch{I}]) =\frac{a_{\stepone, s}[\ch{S}] - a_{\stepone, i}[\ch{I}]}{b_{\stepone, s}[\ch{S}] + b_{\stepone, i} [\ch{I}]+ b_{\stepone, 0}}\,, \\
&\psi_\steptwo([\ch{I}], [\ch{P}])  =\frac{a_{\steptwo, ii}[\ch{I}]^2 +a_{\steptwo, i} [\ch{I}] + a_{\steptwo, ip}[\ch{I}][\ch{P}] + a_{\steptwo, p}[\ch{P}] + a_{\steptwo, pp}[\ch{P}]^2}{b_{\steptwo, ii}[\ch{I}]^2 + b_{\steptwo, i}[\ch{I}] +b_{\steptwo, ip} [\ch{I}][\ch{P}] + b_{\steptwo, p}[\ch{P}] + b_{\steptwo, pp}[\ch{P}]^2 + b_{\steptwo, 0}}\,.
\end{split}
\label{eq_currents_meta}
\end{equation}
Here $\{a_{\bullet,\bullet}\}$ and $\{b_{\bullet,\bullet}\}$ are parameters of the effective model.

The first step to study the energetics of the mechanism in Eq.~\eqref{eq_model_meta} is to write the dynamical systems~\eqref{eq_ds_meta} as in Eq.~\eqref{eq_coarse_grained_dynamics}. We thus introduce the concentration vector and the stoichiometric matrix
\begin{align}
  \boldsymbol{\slow} &= 
  \begin{pmatrix}
    [\ch{S}]	\\
    [\ch{I}] \\
    [\ch{P}] 	
  \end{pmatrix}
  \,,
 &
\cg{\mathbb{S}}= \kbordermatrix{
    & \color{g}\text{1} &\color{g}\text{2}\cr
    \color{g}\ch{S}   & -1  & 0 \cr
    \color{g}\ch{I} & 1  & -1 \cr
    \color{g}\ch{P}   & 0  & 1 \cr
  }\,,
\label{eq_model_meta_network}
\end{align}
as well as the current vector $\boldsymbol \psi(\boldsymbol \slow)=(\psi_\stepone([\ch{S}],  [\ch{I}]), \psi_\steptwo([\ch{I}],  [\ch{P}]))$. We then verify that the dynamical system~\eqref{eq_ds_meta} does not admit cycles: the stoichiometric matrix in Eq.~\eqref{eq_model_meta_network} has no right null eigenvectors. The left null eigenvector of $\cg{\mathbb S}$ is the conservation laws $\boldsymbol \ell = (1,1,1)^\intercal$ corresponding to the total concentration, $L=\boldsymbol \ell\cdot\boldsymbol\slow = [\ch{S}]+[\ch{I}]+[\ch{P}]$. The equilibrium state of Eq.~\eqref{eq_ds_meta} can be identified by solving the system of equations $\boldsymbol \psi(\boldsymbol \slow_{\text{eq}})=0$ and $L=\boldsymbol \ell\cdot\boldsymbol\slow_\text{eq}=\boldsymbol \ell\cdot\boldsymbol\slow(0)$. 
 
\begin{figure}[t]
  \centering
  \includegraphics[width=1\columnwidth]{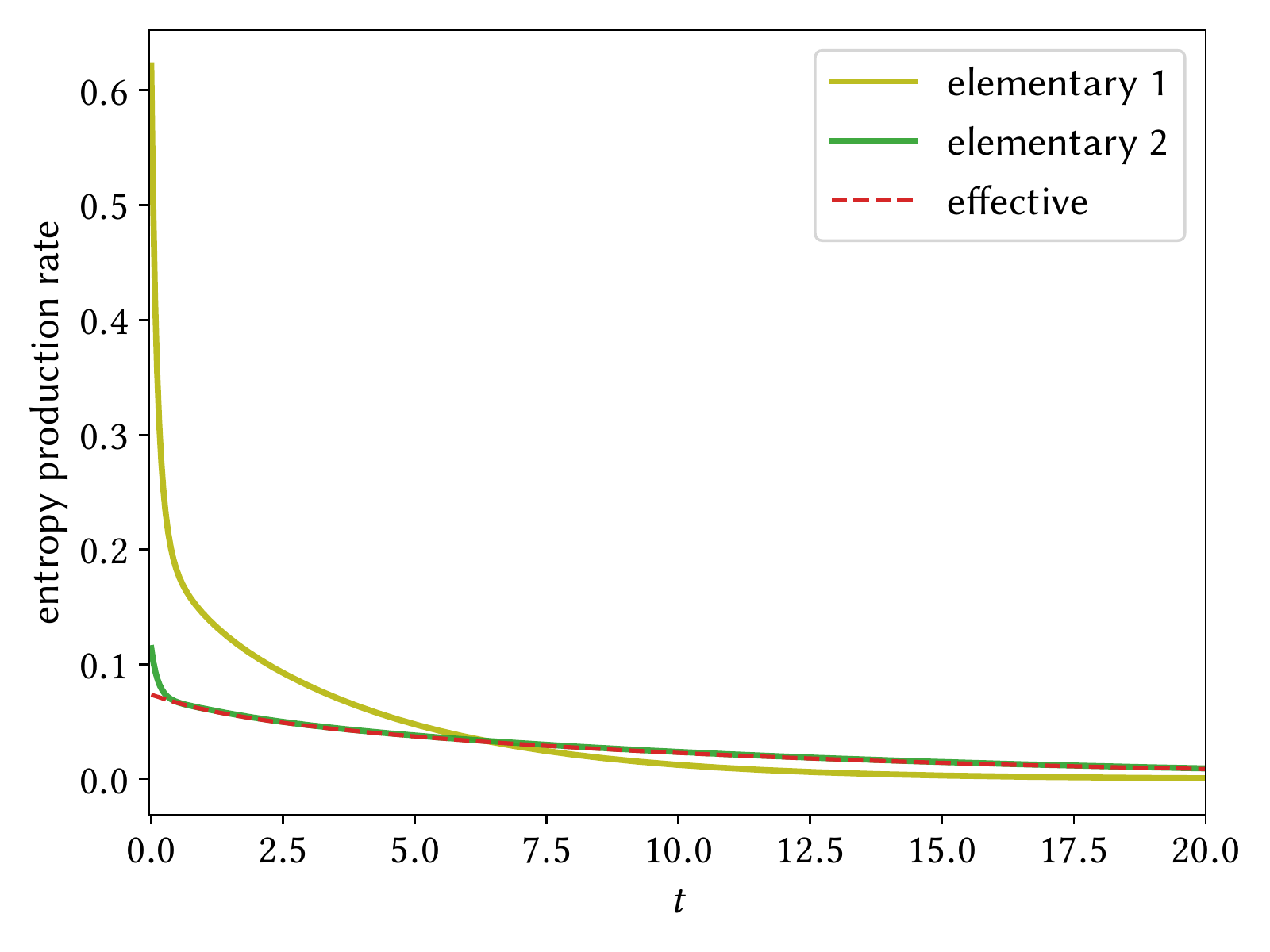}
  \caption{Entropy production rate of the effective mechanism~\eqref{eq_model_meta} (red dashed line), of the elementary mechanism~\eqref{eq_crn_example2} when the \textit{time scale separation hypothesis} does not hold (elementary 1) and when the \textit{time scale separation hypothesis} holds (elementary 2).
The initial concentrations for the slow-evolving species ($[\ch{S}](0)=10 [\ch{I}](0)=10 [\ch{P}](0)=1$) are the same for all the simulations.
The initial concentrations for the fast-evolving species are $[\ch{E_1}](0)=[\ch{E_2}](0)=[\ch{E_3}](0)=10 [\ch{E_1S}](0) =10 [\ch{E_2I}](0)=10 [\ch{E_3I}](0)=0.3$ for elementary 1 and $[\ch{E_1}](0)=0.09$, $[\ch{E_1S}](0) =0.02$, $[\ch{E_2}](0)=[\ch{E_3}](0)=10 [\ch{E_2I}](0)=10 [\ch{E_3I}](0)=0.1$ for elementary 2.
Here, $a_{\stepone,s}=a_{\stepone,i}=0.11$, $b_{\stepone,s}=b_{\stepone,i}=1$, $b_{\stepone, 0}=2$, $a_{\steptwo,ii}=-a_{\steptwo,pp}=0.22$, $a_{\steptwo,i}=-a_{\steptwo,p}=0.44$, $a_{\steptwo,ip}=0$, $b_{\steptwo,ii}=b_{\steptwo,pp}=1$, $b_{\steptwo,i}=b_{\steptwo,p}=b_{\steptwo,0}=4$, $b_{\steptwo,ip}=2$, $\mu^{\circ}_{\ch{S}}=\mu^{\circ}_{\ch{I}}=\mu^{\circ}_{\ch{P}}=\mu^{\circ}_{\ch{E_1}}=\mu^{\circ}_{\ch{E_2}}=\mu^{\circ}_{\ch{E_3}}=1$, $\mu^{\circ}_{\ch{E_1S}}=\mu^{\circ}_{\ch{E_2I}}=\mu^{\circ}_{\ch{E_3I}}=2$, $k_{\pm\elrct}=1$ $\forall \elrct$. For simplicity, we use quantities scaled by some arbitrary reference unit.}
  \label{fig_meta_entropy}
\end{figure}

Assuming that the standard chemical potentials $\mu^{\circ}_{\ch{S}}$, $\mu^{\circ}_{\ch{I}}$ and $\mu^{\circ}_{\ch{P}}$ are known, the second step is to apply the expression of the thermodynamic quantities given in Sec.~\ref{sec_thermo_closed} and~\ref{sec_thermo_open}. For instance, the dissipation can be quantified with the entropy production rate of Eq.~\eqref{eq_cg_entropy_production} which reads now
\begin{equation}
T\cg{\dot{\Sigma}}=(\mu_{\ch{S}} -\mu_{\ch{I}}) \psi_\stepone+(\mu_{\ch{I}} -\mu_{\ch{P}}) \psi_\steptwo\,.
\label{eq_entropy_meta}
\end{equation}
By solving the rate equation~\eqref{eq_ds_meta} for a specific set of parameters and initial condition, one can compute the entropy production rate~\eqref{eq_entropy_meta} which is shown in Fig.~\ref{fig_meta_entropy}. Notice that $\cg{\dot{\Sigma}}(t)\geq 0$ for every time step of the dynamics.
With the same strategy, namely, applying the effective expressions given in this work, one can compute also other thermodynamic quantities.

Consider now the following elementary mechanism
\begin{equation}
\begin{split}
\ch{S + E_1 &<=>[ $k_{+1}$ ][ $k_{-1}$ ] E_1S}\\
\ch{E_1S &<=>[ $k_{+2}$ ][ $k_{-2}$ ] E_1 + I}\\
\ch{I + E_2 &<=>[ $k_{+3}$ ][ $k_{-3}$ ] E_2I}\\
\ch{E_2I &<=>[ $k_{+4}$ ][ $k_{-4}$ ] E_2 + P}\\
\ch{I + E_3 &<=>[ $k_{+5}$ ][ $k_{-5}$ ] E_3I}\\
\ch{E_3I &<=>[ $k_{+6}$ ][ $k_{-6}$ ] E_3 + P}
\end{split}
\label{eq_crn_example2}
\end{equation}
In the limit of fast-evolving enzymes ($\ch{E_1}$, $\ch{E_2}$ and $\ch{E_3}$) and complexes ($\ch{E_1S}$, $\ch{E_2I}$ and $\ch{E_3I}$), the effective dynamics for $[\ch{S}]$, $[\ch{I}]$ and $[\ch{P}]$ provided by the coarse graining procedure discussed in Sec.~\ref{sec_coarse_grained_dynamics} is consistent with the dynamical system~\eqref{eq_ds_meta}. 
We show in Fig.~\ref{fig_meta_entropy} the entropy production rate~\eqref{eq_entropy_production}  of this elementary mechanism for two different set of initial concentrations of the enzymes and the complexes. In the first case, the concentrations are not small enough and the \textit{time scale separation hypothesis} does not hold. Hence, the entropy production rate at the elementary level does not correspond to that of the effective model~\eqref{eq_model_meta}. In the second case, the \textit{time scale separation hypothesis} holds and the entropy production rate at the elementary level is well approximated by the effective one. The initial difference between the two is due to the relaxation of the initial state of the elementary dynamics to the corresponding quasi-steady-state for the fast-evolving species.

\subsection{Example 2}

Consider now the model of gene regulation provided in Ref.~\cite{Feng2014}. It represents the synthesis of two proteins \ch{A} and \ch{B} via the expression of the two genes \ch{G_A} and \ch{G_B}. Each protein promotes its synthesis and represses the synthesis of the other. Then, the proteins degrade. The corresponding chemical reaction network is
\begin{equation}
\begin{split}
\ch{G_A &->[ $\Aexpression$ ][ ] A}\\
\ch{G_B &->[ $\Bexpression$ ][ ] B}\\
\ch{A &->[ $\Adegradation$ ][ ] $\emptyset$}\\
\ch{B &->[ $\Bdegradation$ ][ ] $\emptyset$}\\
\end{split}
\end{equation}
It evolves according to the following dynamical system
\begin{equation}
\left\{
\begin{split}
&\mathrm d_t [\ch{A}](t) = \psi_{\Aexpression}([\ch{A}](t),[\ch{B}](t)) - \psi_{\Adegradation}([\ch{A}](t))\\
&\mathrm d_t [\ch{B}](t) = \psi_{\Bexpression}([\ch{A}](t),[\ch{B}](t)) - \psi_{\Bdegradation}([\ch{B}](t))
\end{split}
\right.
\label{eq_dynamics_gr}
\end{equation}
with the currents
\begin{equation}
\begin{split}
&\psi_{\Aexpression}([\ch{A}],[\ch{B}])= g_{\ch{A}}+ \frac{a_1[\ch{A}]^4}{S^4 + [\ch{A}]^4} + \frac{b_1S^4}{S^4 + [\ch{B}]^4}\,,\\
&\psi_{\Bexpression}([\ch{A}],[\ch{B}])=g_{\ch{B}}+ \frac{a_2[\ch{B}]^4}{S^4 + [\ch{B}]^4} + \frac{b_2S^4}{S^4 + [\ch{A}]^4}\,,\\
&\psi_{\Adegradation}([\ch{A}])=k_{\ch{B}}[\ch{A}]\,,\\
&\psi_{\Bdegradation}([\ch{B}])=k_{\ch{B}}[\ch{B}]\,.\\
\end{split}
\end{equation}
Here, $g_{\ch{A}}$, $g_{\ch{B}}$, $a_1$, $a_2$, $S$, $b_1$, $b_2$, $k_{\ch{A}}$ and $k_{\ch{B}}$ are parameters of the model. The concentration of the proteins \ch{A} and \ch{B} are the only dynamical variables.

This effective model~\eqref{eq_dynamics_gr} is designed in such a way that it cannot equilibrate.
Indeed, the degradation reactions are irreversible and the currents cannot vanish.
The model always relaxes towards a nonequilibrium steady state, but the major issue is that energy sources preventing the equilibration cannot be accounted for.
While our thermodynamic quantities can be formally defined for this model (they just require the dynamical system and the standard chemical potential), they are meaningless. 

We note that the use of irreversible reactions and/or nonvanishing currents is a very common feature of kinetic models for biology and does not preclude \textit{per se} a consistent thermodynamic analysis as recently illustrated for the irreversible Michaelis-Menten enzymatic scheme~\cite{Voorsluijs2020}.


\section{Conclusions\label{sec_conclusions}}
In this work, we developed a thermodynamic theory for effective (non-mass-action) models of both closed and open \crn s.
We focused here only on deterministic models.
Our approach provides the exact thermodynamic quantities when the effective models result from underlying elementary (mass-action) networks satisfying the \textit{time scale separation hypothesis}.
This was proven by exploiting the topological properties of the \crn s.
Our  \textit{time scale separation hypothesis} can be considered as a zero-order expansion in the ratio between the fast and the slow time scale.
Exploring higher order corrections and whether they can be used to bound deviations in entropy production rates is left for future work.

Similar approaches might be employed in other frameworks.
First, the topological properties could be used to derive a thermodynamically consistent coarse-graining of stochastic \crn s.
Second, one can exploit weakly broken conservation laws to collect different species into effective mesostates.
These mesostates may then satisfy closed evolution equations.
For example, during a catalytic process the complex enzyme-substrate is transformed in many different species that can be considered as a unique mesostate using the conservation of the total concentration of the enzyme.
We leave also these points to future investigations.

\section{Acknowledgments}
This research was funded by the European Research Council project NanoThermo (ERC-2015-CoG Agreement No.~681456).



\appendix
\section{Reference chemical potentials\label{app_reference_chemical_potential}}
We discuss here the constraints between the chemical potentials at equilibrium. We consider the case of the effective dynamics, but the same exact reasoning applies to the elementary dynamics (see, for example, Appendix A in Ref.~\cite{Avanzini2019a}). Because of the local detailed balance~\eqref{eq_cg_local_detailed_balance}, the vector of the equilibrium chemical potentials $\cg{\boldsymbol\mu}_{\text{eq}}$ is a left-null eigenvectors of $\cg{\mathbb S}$ and, therefore, it can be expressed as a linear combination of the conservation laws $\cg{\boldsymbol l}{}^{\slowconslaw}$:
\begin{equation}
\cg{\boldsymbol\mu}_{\text{eq}}^\intercal = \boldsymbol{ \coefficient} \cdot \cg{\matrixconslaw}\,,\label{eq_app_eqchempotential_tot}
\end{equation}
where we rewrite Eq.~\eqref{eq_chem_potential_eq_cons_law} by introducing the vector $\boldsymbol{ \coefficient} = (\dots,\coefficient_{\slowconslaw},\dots)^\intercal$ and the matrix $\cg{\matrixconslaw}$ with entries $\{\ell{}^{\slowconslaw}_{\chemspecies}\}_{\chemspecies\in \slowset}$. 

We examine the constraints imposed on the chemical potentials by Eq.~\eqref{eq_app_eqchempotential_tot} as if the \crn\text{ } were open. The chemical species  are either internal species $\internalset$, or chemostatted species $\chemostattedset_\breaker$ breaking the conservation laws or other chemostatted species $\chemostattedset_\nobreaker$. The set of conservation laws splits into unbroken conservation laws $\{\cg{\boldsymbol\ell}{}^\unbrokenconslaw\}$ and broken conservation laws $\{\cg{\boldsymbol\ell}{}^\brokenconslaw\}$. We show that the equilibrium chemical potentials of the $\chemostattedset_\breaker$ species set the equilibrium chemical potentials of the $\chemostattedset_\nobreaker$ species. We start applying the same splitting to $\cg{\matrixconslaw}$ 
\begin{equation}
\cg{\matrixconslaw}=
\renewcommand*{\arraystretch}{1.5}
 \kbordermatrix{
    & \color{g}\internalset &\color{g}&\color{g}\chemostattedset_\nobreaker&\color{g}\chemostattedset_\breaker\cr
    \color{g}\unbrokenconslaw    &\matrixconslawXun &\color{g}\vrule&\mathbb 0 &\mathbb 0\cr
    \color{g}\brokenconslaw   &\matrixconslawXbr &\color{g}\vrule&\matrixconslawforces &\matrixmoieties\cr
  }\,,
  \label{eq_matrixconslaw}
\end{equation}
and $\boldsymbol{ \coefficient} =( \boldsymbol{ \coefficient}_{\text{un}} , \boldsymbol{ \coefficient}_{\text{br}} )$. Here, we introduced the vectors $ \boldsymbol{ \coefficient}_{\text{un}} = (\dots,\coefficient_{\unbrokenconslaw},\dots)^\intercal$ and $\boldsymbol{ \coefficient}_{\text{br}}=(\dots, \coefficient_{\brokenconslaw},\dots)^\intercal$, and the matrices $\matrixconslawXun$ with entries $\{\ell{}^{\unbrokenconslaw}_{\chemspecies}\}_{\chemspecies\in \internalset}$, $\matrixconslawXbr$ with entries $\{\ell{}^{\brokenconslaw}_{\chemspecies}\}_{\chemspecies\in \internalset}$ and $\matrixconslawforces$  with entries $\{\ell{}^{\brokenconslaw}_{\chemspecies}\}_{\chemspecies\in \chemostattedset_\nobreaker}$. The matrix $\matrixmoieties$ was already introduced in Subs.~\ref{subs_broken_conservation_laws}. The zero matrices $\mathbb 0$ collect the entries of the unbroken conservation laws for chemostatted species (which vanish by definition).

Using Eq.~\eqref{eq_app_eqchempotential_tot} and~\eqref{eq_matrixconslaw}, we can recognize that the equilibrium chemical potentials of the $\chemostattedset_\breaker$ and  $\chemostattedset_\nobreaker$ species are given by 
\begin{equation}
\cg{\boldsymbol\mu}_{\chemostattedset_\breaker, \text{eq}}^\intercal = \boldsymbol{ \coefficient}_{\text{br}} \cdot \matrixmoieties\,,
\end{equation}
\begin{equation}
\cg{\boldsymbol\mu}_{\chemostattedset_\nobreaker, \text{eq}}^\intercal = \boldsymbol{ \coefficient}_{\text{br}} \cdot \matrixconslawforces\,,\label{eq_app_chempotential_nobreaker}
\end{equation}
respectively. The matrix  $\matrixmoieties$ is square and nonsingular, and it can be inverted. Hence
\begin{equation}
\boldsymbol{ \coefficient}_{\text{br}}^\intercal=\cg{\boldsymbol\mu}_{\chemostattedset_\breaker, \text{eq}} \cdot \matrixmoieties^{-1}\,.\label{eq_app_coefficient_broken}
\end{equation}
The equilibrium chemical potentials of the $\chemostattedset_\nobreaker$ species thus become
\begin{equation}
\cg{\boldsymbol\mu}_{\chemostattedset_\nobreaker, \text{eq}}^\intercal =\cg{\boldsymbol\mu}_{\chemostattedset_\breaker, \text{eq}} \cdot \matrixmoieties^{-1}\matrixconslawforces\,,
\end{equation}
proving that they depend on the equilibrium chemical potentials of the $\chemostattedset_\breaker$ species.

In open \crn s, the chemostats fix the chemical potentials of the $\chemostattedset$ species and, so their concentrations. The $\chemostattedset_\breaker$ chemostats set a reference equilibrium conditions, meaning 
\begin{equation}
\cg{\boldsymbol\mu}_{\chemostattedset_\breaker, \text{eq}} = {\boldsymbol\mu}_{\chemostattedset_\breaker}\,\label{eq_app_eqchempotential_potential}
\end{equation}
\begin{equation}
\cg{\boldsymbol\mu}_{\chemostattedset_\nobreaker, \text{eq}}^\intercal ={\boldsymbol\mu}_{\chemostattedset_\breaker} \cdot \matrixmoieties^{-1}\matrixconslawforces\,.\label{eq_app_eqchempotential_force}
\end{equation}
Here ${\boldsymbol\mu}_{\chemostattedset_\breaker}$ is the vector of chemical potentials set by the $\chemostattedset_\breaker$ chemostats. It is the same whether we consider the elementary dynamics or the effective one. The $\chemostattedset_\nobreaker$ chemostats impose the chemical potentials ${\boldsymbol\mu}_{\chemostattedset_\nobreaker}$ which, in general, do not satisfy Eq.~\eqref{eq_app_eqchempotential_force}. As a consequence, the nonconservative forces
\begin{equation}
\boldsymbol{\mathcal F}=(\dots, \mu_\chemspecies-(\boldsymbol\mu_{\refchempot}\cdot\matrixmoieties^{-1})_{\brokenconslaw}l^{\brokenconslaw}_\chemspecies, \dots)_{\chemspecies\in Y}^\intercal \label{eq_app_nonconservative_force}
\end{equation}
of Eq.~\eqref{eq_nc_work} are generated. Indeed, Eq.~\eqref{eq_app_nonconservative_force} can be rewritten as
\begin{equation}
\boldsymbol{\mathcal F}=(\dots, \mu_\chemspecies-\mu_{\chemspecies,\text{eq}}, \dots)_{\chemspecies\in Y}^\intercal\,,
\end{equation}
with $\mu_{\chemspecies,\text{eq}}$ given in Eq.~\eqref{eq_app_eqchempotential_potential} and~\eqref{eq_app_eqchempotential_force}. Each entries in $\boldsymbol{\mathcal F}$ represents the difference between the chemical potential set by a chemostat and the equilibrium one. Therefore, it represents the force keeping the system out of equilibrium.

Note that, thanks of Eq.~\eqref{eq_app_eqchempotential_potential}, we can write $\boldsymbol{ \coefficient}_{\text{br}}^\intercal={\boldsymbol\mu}_{\chemostattedset_\breaker} \cdot \matrixmoieties^{-1}$. We use this property and the definition of the moieties~\eqref{eq_cg_moieties} in the last step of Eq.~\eqref{eq_finalstep_semigran_equilibrium}.

%
%

\bibliography{biblio}

\end{document}